\documentclass{aa}
\usepackage{mathrsfs}
\usepackage{lscape}
\usepackage{graphicx}
\usepackage{txfonts}
\def\kms{\hbox{km\,s$^{-1}$}}

\newcommand{\teff}{log\,T$_{\rm eff}\,$[K]}
\newcommand{\lpr}{${\mathscr L}/{\mathscr L}_\odot$}

\newcommand{\vsini}{$v\,$sin$\,i$}

\begin{document}

   \title{Mapping the core of the Tarantula Nebula with VLT-MUSE\thanks{Based on observations made with ESO telescopes
at the Paranal observatory under programme ID 60.A-9351(A).} }
   \subtitle{II. The spectroscopic Hertzsprung-Russell diagram of OB stars in NGC\,2070 }
   \author{N.~Castro\inst{1}, P.~A.~Crowther\inst{2}, C.~J.~Evans\inst{3}, J.~S.~Vink\inst{4}, J. Puls\inst{5}, A. Herrero\inst{6,7}, M. Garcia\inst{8}, F. J. Selman\inst{9}, M.~M. Roth\inst{1} and  S. Sim\'on-D\'iaz\inst{6,7}}
     \institute{Leibniz-Institut für Astrophysik Potsdam (AIP), An der Sternwarte 16, 14482 Potsdam, Germany\\
              \email{ncastro@aip.de}
          \and
            Department of Physics \& Astronomy, University of Sheffield, Hounsfield Road, Sheffield, S3 7RH, UK
         \and
                  UK Astronomy Technology Centre, Royal Observatory, Blackford Hill, Edinburgh, EH9 3HJ, UK
                                   \and
                  Armagh Observatory and Planetarium, College Hill, Armagh BT61 9DG, Northern Ireland, UK
                  \and
                  LMU M\"unchen, Universit\"atssternwarte Scheinerstr. 1, D-81679 M\"unchen, Germany
                  \and
                  Instituto de Astrof\'isica de Canarias, La Laguna, Tenerife, Spain
                  \and
                  Departamento de Astrof\'isica, Universidad de La Laguna, Tenerife, Spain
                  \and
                  Centro de Astrobiolog\'ia CSIC-INTA. Crtra. de Torrej\'on a Ajalvir km 4. E-28850 Torrej\'on de Ardoz, Madrid, Spain
                  \and
                  European Southern Observatory, Alonso de Cordova 3107 Vitacura Casilla, 7630355, Santiago, Chile
 } 
\titlerunning{The sHRD of OB stars in NGC\,2070} \authorrunning{N. Castro et al.} 

\abstract {We present the spectroscopic analysis of 333 OB-type stars extracted from
  VLT-MUSE observations of the central 30\,$\times$\,30\,pc of
  NGC\,2070 in the Tarantula Nebula on the Large Magellanic Cloud,
   the majority of which are analysed for the the first time. The
  distribution of stars in the spectroscopic Hertzsprung-Russell
  diagram (sHRD) shows 281 stars in the main sequence. We find two groups in
  the main sequence, with estimated ages of 2.1$\pm$0.8 and
  6.2$\pm$2\,Myr. A subgroup of 52 stars is apparently beyond
  the main sequence phase, which we consider to be due to emission-type objects
   and/or significant nebular contamination affecting the analysis. As in  previous
   studies, stellar masses derived from the sHRD are
  systematically larger than those obtained from the conventional HRD,
  with the differences being largest for the most massive stars.
  Additionally, we do not find any trend between the estimated
  projected rotational velocity and evolution in the sHRD. The
  projected rotational velocity distribution presents a tail of fast
  rotators that resembles findings in the wider population of
  30~Doradus. We use published spectral types to calibrate the
  \ion{He}{i}$\lambda4921$/\ion{He}{ii}$\lambda5411$ equivalent-width
  ratio as a classification diagnostic for early-type main sequence
  stars when the classical blue-visible region is not observed.
  Our model-atmosphere analyses demonstrate that the resulting
  calibration is well correlated with effective temperature.}

   \maketitle 

\section{Introduction}

Massive stars are thought to drive the chemical and dynamical
evolution of galaxies \citep{2009ApJ...695..292C,2012ARA&A..50..531K}.
They are also thought to be strong candidates for the re-ionisation of the early Universe
\citep[e.g.][]{2010Natur.468...49R}. However, the formation and
evolution of stars more massive than 10\,M$_\odot$ still hold many
unanswered questions \citep{2012ARA&A..50..107L}.  These uncertainties
quickly grow as we move up to higher masses in the
Hertzsprung-Russell diagram (HRD), particularly for the most massive
stars (with $M$\,$>$\,100\,$M_\odot$) and beyond the hydrogen
core-burning phase \citep{2015HiA....16...51V}.

The HRD is a powerful tool for investigating the evolution of massive
stars, particularly the influence that factors such as mass, rotation,
metallicity, magnetic fields, and binarity have on their main sequence
lifetimes, later evolutionary phases, and ultimate fates
\citep[e.g.][]{1981ARA&A..19..277S,1987A&A...178..159M,2000ARA&A..38..143M,2014ApJ...782....7D}.
Disentangling the respective roles of these factors to chart the
various evolutionary paths requires homogeneous, statistically robust
studies of the physical properties of populations of massive stars.

Historically, photometric studies were the sole route for investigating large
stellar samples
\citep[e.g.][]{1990ApJ...363..119F,2002ApJS..141...81M}, but the
degeneracy of the optical colours to derive the effective temperatures (T$_{\rm eff}$) of OB-type
stars  \citep{1988ApJ...328..704H}  from such an approach limits
any insight into stellar evolution
\citep{2011A&A...532A.147L}. More recently, spectroscopic surveys have
transformed the field, enabling observations of large samples for a detailed
quantitative study in the Milky Way
\citep[e.g.][]{2017A&A...597A..22S,2017A&A...599A..30M},
the Magellanic Clouds
\citep[e.g.][]{2004MNRAS.353..601E,2011A&A...530A.108E,2004ApJ...608.1001M,2018A&A...615A..40R,2019A&A...625A.104R,2019arXiv190503359D},
and in nearby star-forming galaxies in the Local Group
\citep[e.g.][]{2003ApJ...584L..73U,2012A&A...542A..79C}.

Based on this wealth of spectroscopic data,
the spectroscopic HRD \citep[sHRD; ${\mathscr L}$\,$\equiv$\,T$_{\rm
  eff}^4/g$;][]{2014A&A...564A..52L} can provide insights
into stellar evolution \citep[see also][]{2014A&A...570L..13C}. The sHRD,
 which is the inverse of the flux-weighted gravity introduced by
 \cite{2003ApJ...582L..83K},  does not require knowledge
of the extinction or distance to the targets and can be calculated
from the stellar analyses. In contrast to the
classical Kiel diagram (T$_{\rm eff}$\,--\,log\,$g$), the sHRD sorts stars
according to their proximity to the Eddington limit.
\cite{2014A&A...570L..13C,2018ApJ...868...57C}  proposed empirical anchors for the stellar
evolution of massive stars in the Milky Way and the Small Magellanic Cloud (SMC) based on the sHRD, such as the
position of the zero age main sequence (ZAMS) and the terminal age
 main sequence (TAMS). These provide robust targets for theories of
massive-star evolution, including constraints on parameters such as
rotation, convective overshooting, and metallicity
\citep{2011A&A...530A.115B,2012A&A...537A.146E,2017A&A...597A..71S,2019A&A...622A..50H}.
In this study, we extend this approach to the massive-star population
of the Large Magellanic Cloud (LMC), which has an intermediate metallicity ($Z/Z_\odot=0.5$)
between that of the Milky Way and the SMC.

Although conceptually designed as a cosmology machine, the Multi-Unit Spectroscopic Explorer
 \citep[MUSE;][]{2014Msngr.157...13B} on the Very Large Telescope (VLT) presents
  exciting capabilities for the spectroscopy of stellar populations
 \citep[see the review by][]{2019AN....340..989R} and a leap forward from pioneering studies
based on 3D spectroscopy \citep[e.g.][]{2013A&A...549A..71K}. This unique integral field
 spectrograph with a large field-of-view (1$'$$\times$1$'$),
excellent image quality, and high efficiency enables a novel approach to studying populations of
massive stars, somewhere in between traditional photometric and multi-object
spectroscopic surveys.   MUSE also overcomes selection biases caused by extinction.
Internal extinction may be significant within star-forming galaxies, and O stars may be missed
in surveys where targets were selected based on optical colours only \citep[e.g.][]{2019MNRAS.484..422G}.
MUSE capabilities in the analysis of  young stellar populations have been explored
in recent years in nearby clusters \citep[e.g.][]{2015A&A...582A.114W,2018AJ....156..211Z},
the Magellanic Clouds \citep[e.g.][]{2019MNRAS.486.5263M,2020A&A...634A..51B}, and, even
farther away in galaxies outside the Local Group \citep[e.g. NGC300 and Leo P;][]{2018A&A...618A...3R,2019A&A...622A.129E}.

The Tarantula Nebula in the LMC is the most
luminous star-forming complex in the Local Group
\citep{1984ApJ...287..116K,2019Galax...7...88C}. The inner part of the Tarantula,
NGC\,2070, hosts a well-known rich population of OB-type and
Wolf-Rayet (W-R) stars
\citep[e.g.][]{1985A&A...153..235M,1999A&A...341...98S,2011A&A...530A.108E}.
Moreover, in the core of NGC\,2070 lies the young massive cluster R136,
 home to the most massive stars known to date
\citep{2010MNRAS.408..731C,2020MNRAS.499.1918B}.

To test the unique capabilities of MUSE, NGC\,2070 was observed as part
of its Science Verification (SV) programme. The SV observations have
provided the most complete spectroscopic census of the region to date
\citep[][hereafter Paper I]{2018arXiv180201597C}. Here we apply a similar approach to that
of \cite{2014A&A...570L..13C,2018ApJ...868...57C} to the MUSE data to
investigate massive-star evolution at the metallicity of the LMC (and
notably in a young region that is still undergoing active
star formation). A big advantage of NGC\,2070 compared to the samples
in the Milky Way and the SMC
\citep{2014A&A...570L..13C,2018ApJ...868...57C} is that it is a more
homogeneous population in terms of age, facilitating the study of the properties of the most massive stars.
 Large and homogeneous studies are crucial to reducing
 the biases outlined by  \cite{2014A&A...570L..13C} and providing reliable empirical boundaries for theoretical models of stellar evolution.

This article is structured as follows. Section~\ref{Sect:Data} briefly
introduces the data and the sample for quantitative analysis.
Section~\ref{Sect:Analisis} describes the methods used to estimate
physical parameters and to classify the spectra. Section~\ref{Sect:shrd}
introduces the sHRD for NGC\,2070, and Sect.~\ref{discussion}
discusses the ages, masses, and rotation rates of the sample. We
summarise our findings in Sect.~\ref{Sect:Summ}.

\section{Data and sample selection}
\label{Sect:Data}

The central 2$'$\,$\times$\,2$'$ of NGC\,2070 was observed with a
mosaic of four MUSE pointings as part of the SV programme in August
2014. The data were obtained with the natural seeing mode (with
spatial sampling of 0\farcs2 on the sky) and using the extended
wavelength coverage, providing spectra over 4595-9365\,\AA, at a
resolving power of $R$\,$\sim$3000 around H$\alpha$.

The observations, data reduction, and a comprehensive catalogue of
early-type stars in the MUSE mosaic were presented in Paper I. For the quantitative analysis presented
here, we used spectra extracted from the 4$\times$600\,s exposures
and limited our census to OB-type spectra with signal-to-noise
ratios (S/N) greater than 50. This enabled a robust study of their
physical properties from comparison with synthetic spectra from model
atmospheres.

\begin{figure}[]
        \resizebox{\hsize}{!}{\includegraphics[angle=0,width=\textwidth]{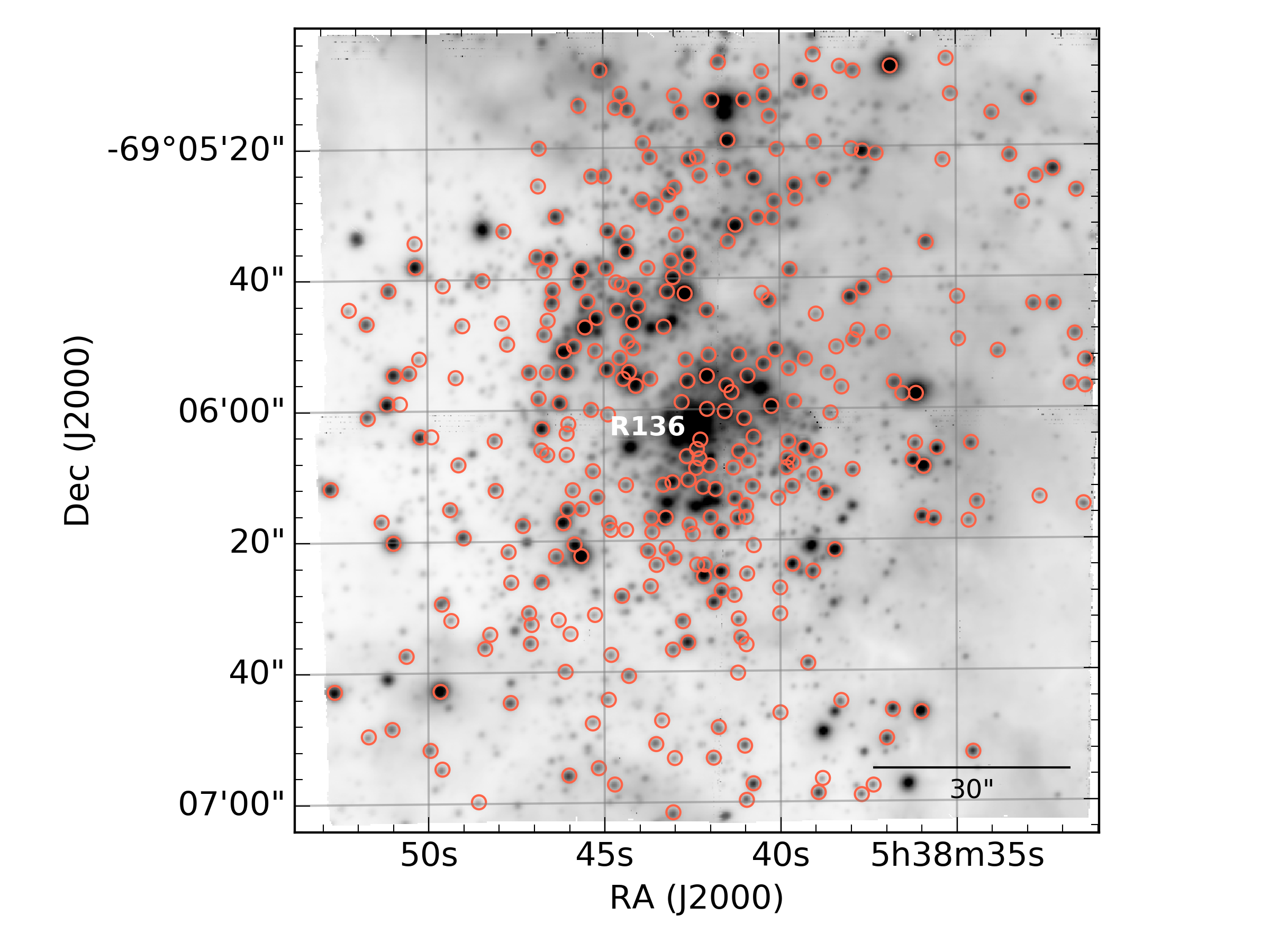}}
        \caption{Spatial distribution of the sample stars (open red circles)
          overlaid on the continuum-integrated 2$'$\,$\times$2\,$'$ MUSE
          mosaic in NGC\,2070 (see Paper I). The core of the cluster, R136, is marked.}\label{Fig:FoV}
\end{figure}

We excluded stars with T$_{\rm eff}$\,$<$\,10\,000\,K as well as W-R stars
(and related stars) with \ion{He}{ii}\,$\lambda$5411 in emission. Such
objects are not covered by our current grid of models (see
Sect.~\ref{Sect:Analisis}) and will be analysed in the future with the
appropriate tools. In total, we considered 333 OB-type spectra with
S/N\,$>$\,50. Their spatial distribution in the MUSE mosaic is shown
in Fig.~\ref{Fig:FoV}, and their positions are listed in
Table~\ref{TAB:cat}.

Each spectrum was visually examined to discard composite spectra due to project or unresolved
stars or binaries. Nevertheless, unresolved components contributing to the spectra must be
kept in mind in the conclusions. The R136 cluster at the core of NGC\,2070 is unresolved and is
not included in this study (Fig.~\ref{Fig:FoV}). A parallel project using the narrow-field
mode  \citep{2019Msngr.176...16L} of MUSE to study R136 with superior spatial
resolution is ongoing.

\section{Spectral analysis}
\label{Sect:Analisis}

\begin{figure}
        \centering
        \includegraphics[width=9.5cm]{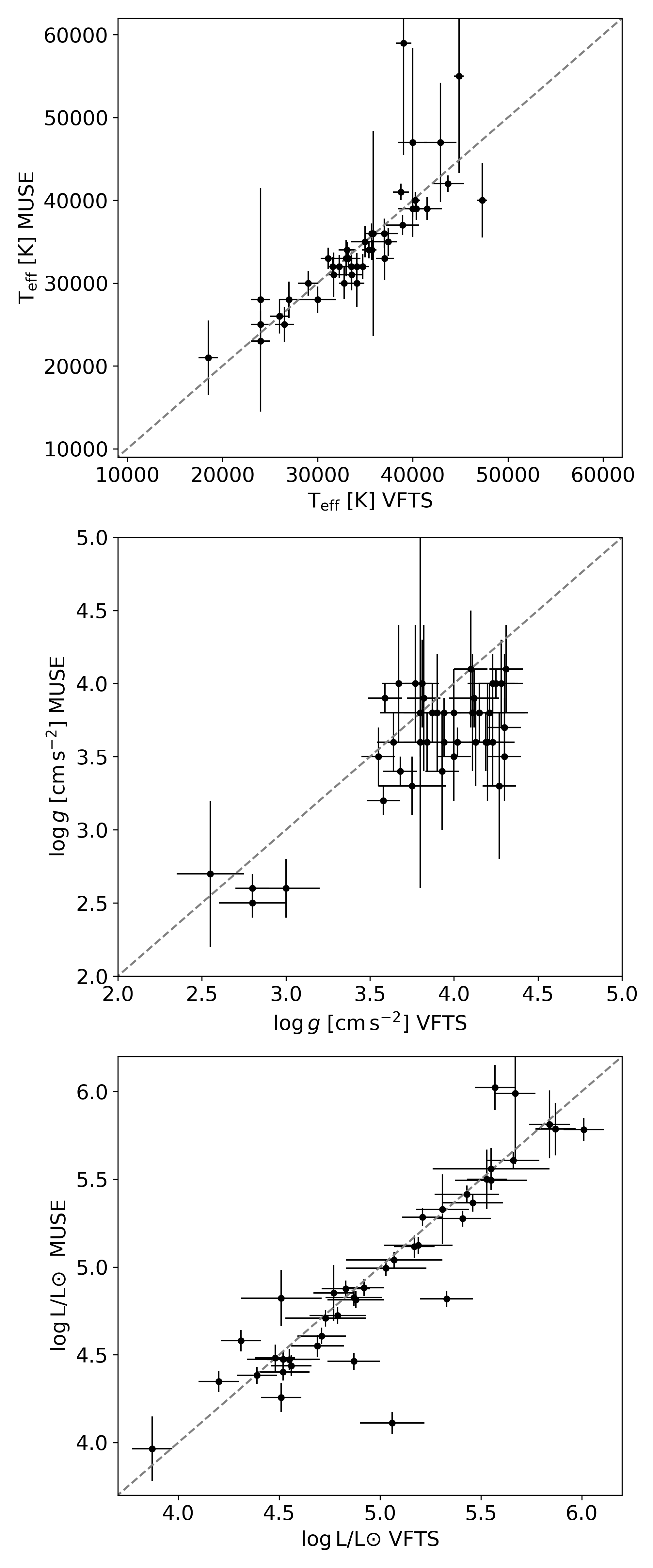}
        \caption{Estimated effective temperatures (T$_{\rm eff}$,
          top panel), gravities (log\,$g$, middle panel), and luminosities (log\,L/L\sun, bottom panel) from the MUSE data
          compared with results from the VFTS for the 42 stars in common.          \citep{2014A&A...564A..39S,2017A&A...601A..79S,2015A&A...575A..70M,2017A&A...600A..81R,2017A&A...603A..91G,2018Sci...359...69S}. }\label{Fig:VFTS}
\end{figure}

The MUSE wavelength range does not include the traditional diagnostic
features (e.g. \ion{He}{i}\,$\lambda$4471, \ion{He}{ii}\,$\lambda$4542,
\ion{and Si}{iii}\,$\lambda$4552) used in the spectral classification and
quantitative analysis of OB-type stars
\citep[e.g.][]{1990PASP..102..379W}.  Nevertheless, the MUSE range
includes lines from both \ion{He}{i} ($\lambda\lambda$4713, 4921,
5876, 6678) and \ion{He}{ii} ($\lambda\lambda$5411, 6683), which can be
used to constrain stellar parameters (\citealt{2017Msngr.X...XC} and Paper I). Both H$\alpha$ and
H$\beta$ are included in the MUSE spectra, but in a region as young as
NGC\,2070, these are generally contaminated by strong nebular emission
from the surrounding excited gas.

Our sample was analysed using a grid of hydrogen and helium (HHe) {\sc fastwind} models
\citep{1997A&A...323..488S,2005A&A...435..669P,2012A&A...543A..95R}.
{\sc fastwind} calculates the atmosphere and line formation in
spherical symmetry, with an explicit treatment of the stellar wind and
taking into account non-local thermodynamic equilibrium.  The analysis
used a grid similar to the one described by \cite{2018ApJ...868...57C}
but now calculated with the LMC metallicity (i.e.  $Z/Z_\odot=0.5$). The grid temperatures and gravities range from 9\,000\,$<$\,T$_{\rm eff}$\,$<$\,67\,000\,K (in steps
of 1000\,K) and 0.8\,$<$\,log\,$g$\,$<$\,5.0\,dex (in steps of
0.1\,dex). The grid was built with a low wind strength parameter, log$\,Q=-14$
\citep{1996A&A...305..171P,2000A&AS..141...23P}. In the upper part of the HRD, stronger stellar winds
are expected to affect lines such as HeII$\lambda4686$; however, this line is  not included in the
analysis. Helium abundances (Y$_{He}$)
were set to $0.09$.  \cite{2018ApJ...868...57C}  tested different surface helium abundances and
concluded that it was not possible to constrain the He abundance with the modest spectral resolution.

The analysis used a similar $\chi^2$ algorithm to the one used by
\cite{2012A&A...542A..79C,2018ApJ...868...57C}. The algorithm
simultaneously fits the available He lines and, where possible, the
wings of the H$\beta$ profile for each star and searches for the
set of parameters that best reproduces the observations \citep[see
also][]{2005ApJ...622..862U,2007ApJ...659.1198E}. H$\beta$ is included
to try to constrain the surface gravity, but we note that it is influenced
in many instances by the strong nebular contribution and/or strong stellar
wings for the most massive stars \citep{2018ApJ...868...57C}. Comparisons with previous studies in the LMC (see Sect.~\ref{sect:vfts}) reinforce
the fact that the reader should be cautious of the quoted gravities, but they
serve as a first estimate to investigate the overall sHRD of the
region.

Projected rotational velocities (\vsini) were estimated by fitting the
He lines with the {\sc fastwind} model from the first iteration of the
fitting algorithm, where the model is broadened assuming only
contributions from stellar rotation and the instrumental resolution.
The atmospheric analysis was then repeated with the new \vsini\
estimate until convergence. Macroturbulence is expected to be an
important additional broadening factor in some stars in the upper part
of the sHRD \citep[see][]{2016A&A...593A..14G,2017A&A...597A..22S}.
However, given the limitations of the modest spectral resolution, we do not include
this in our analysis, with the result that some of the quoted \vsini\
values may be overestimated.

The ionisation balance of \ion{He}{ii}/\ion{He}{i} is used to estimate
T$_{\rm eff}$ for spectra where both species are available. However,
in the earliest O-type stars, we lack suitable \ion{He}{i} lines, and
\ion{He}{ii} is absent for all but the earliest B-types. For these
parts of the sample, our analysis rests on the accuracy of the HHe
models, and some degeneracy is to be expected \cite[and is indirectly
included in the estimated errors from the algorithm; see Sect.~3
of][]{2018ApJ...868...57C}. Nebular contamination in the helium lines may be expected after
the sky subtraction, also affecting the effective temperatures derived in our analysis.
The accuracy of the temperatures from our pure HHe analysis is discussed in the next
sections  \citep[see also the discussion in][]{2018ApJ...868...57C}.
The parameters derived from this work are summarised in Table~\ref{TAB:cat}.

\subsection{Comparisons with VLT-FLAMES Tarantula Survey}
\label{sect:vfts}

We cross-matched our sample with stars observed as part of the
VLT-FLAMES Tarantula Survey \cite[VFTS;][]{2011A&A...530A.108E}.
Excluding stars with \teff\,$<$\,4.3 from our analysis (identified as
possible emission-line stars with unreliable temperatures from our
approach; see Sect.~\ref{Sect:post}), there are 42 stars in common.

Table~\ref{TAB:Vfts} compares our results with those obtained from the
relevant VFTS studies
\citep{2014A&A...564A..39S,2017A&A...601A..79S,2015A&A...575A..70M,2017A&A...600A..81R,2017A&A...603A..91G,2018Sci...359...69S}.
The VFTS stellar parameters were also determined using standard
techniques and the {\sc fastwind} stellar atmosphere
code\footnote{With the exception of the study from \cite{2017A&A...603A..91G},
who used {\sc tlusty} model atmospheres \citep{1995ApJ...439..875H}.}. Within
the uncertainties, there is reasonably good agreement between the estimated
temperatures (Fig.~\ref{Fig:VFTS}), albeit with some significant outliers at
the highest values. We find a good match, with some outliers, between the
luminosities estimated in this work (see Sect.~\ref{Sect:Masa}) and VFTS results as well.

\begin{figure*}[]
        \resizebox{\hsize}{!}{\includegraphics[angle=0,width=\textwidth]{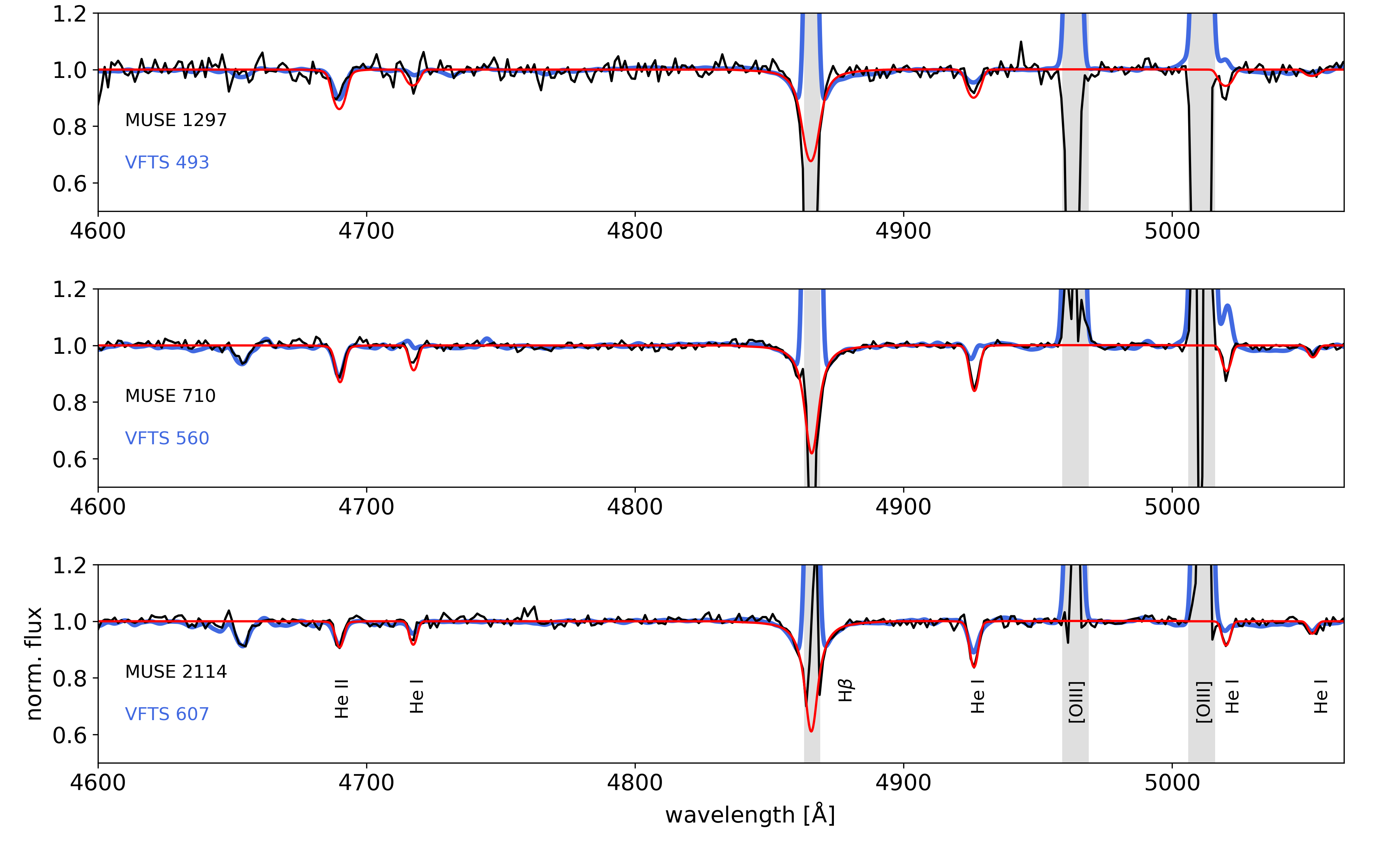}}
        \caption{Comparison between three stars analysed in this work (solid black line)
                and in the VFTS project (solid blue line) whose derived gravities differ significantly between studies. The best {\sc fastwind} model obtained in the MUSE
                analysis is shown (solid red lines). Wavelengths expected to be strongly affected
                by nebular contamination are marked in grey. }\label{Fig:VFTS_EXA}
\end{figure*}

However, there is a large scatter between the estimated surface gravities (Fig.~\ref{Fig:VFTS}), with the MUSE estimates
systematically $\sim$0.3\,dex lower than those from the VFTS results.
This spread in estimated gravities is not unexpected given the limited
diagnostics available and the strong contamination by nebular
emission. Figure~\ref{Fig:VFTS_EXA} compares the spectra of three stars whose  MUSE and VFTS analyses differ significantly and which have strong nebular contamination. MUSE\,1297 (VFTS\,493)
shows a remarkably large discrepancy in the gravity ($\sim 1\,$dex, Table~\ref{TAB:Vfts})
compared to previous VFTS analyses. However,  our algorithm also provides large errors, highlighting the limitations of modelling this O8-9 dwarf.  An over-normalisation of the H$\beta$ continuum,
due to the spectral resolution and strong nebular contamination, could be behind the
discrepancy. We analysed the three VFTS stars in Fig.~\ref{Fig:VFTS_EXA} using the grid
described in this work and the blue wavelengths available in the VFTS spectra, that is, including
the $\sim4000-5000$\,\AA\ region not covered by MUSE. We got larger gravities,
drastically reducing the large differences in these three stars (see Table~\ref{TAB:Vfts})
by $\sim0.4\,$dex. Additional spectroscopic anchors observed both at MUSE and at bluer optical
wavelengths are critical for better constraining the gravity and reducing the errors.
This should be borne in mind during the discussion of the sHRD (Fig.~\ref{Fig:SAMPLE}) and the
mass discrepancies discussed in Sect.~\ref{Sect:Masa}.

\begin{table*}
        \centering
        \caption{ Estimated effective temperatures, surface
gravities, and luminosities from MUSE (cf. published values from the VFTS).}       \label{TAB:Vfts}
\begin{tabular}{llcccccccccccc}
\hline\hline
& & \multicolumn{2}{c}{MUSE} & \multicolumn{2}{c}{VFTS}
& \multicolumn{2}{c}{MUSE} & \multicolumn{2}{c}{VFTS} & \multicolumn{2}{c}{MUSE} & \multicolumn{2}{c}{VFTS}\\
& & T$_{\rm eff}$ & $\sigma_{T_{\rm eff} }$ & T$_{\rm eff}$ &  $\sigma_{T_{\rm eff} }$ &
log\,$g$ & $\sigma_{log\,g}$ & log\,$g$ &$\sigma_{log\,g}$ & log\,L& $\sigma_{log\,L}$& log\,L &$\sigma_{log\,L}$ \\
MUSE ID & VFTS ID & (K) & (K)  & (K) & (K) & (dex) & (dex) & (dex) & (dex)  &[$L_\odot$]  & [$L_\odot$] & [$L_\odot$] & [$L_\odot$] \\
\hline
830 & 465 &  59000 & 13500 & 39050 & 820 & 4.0 & 0.4 & 3.77 & 0.10  & 6.023 & 0.127 & 5.57 & 0.10 \\
1068 & 518 & 55000 & 11700 & 44850 & 500 & 4.0 & 0.4 & 3.67 & 0.10 & 5.989 & 0.403 & 5.67 & 0.10 \\
2901 & 382 & 47000 & 11400 & 40000 & 1500 & 4.0 & 0.3 & 3.81 & 0.10& 5.329 & 0.199 & 5.31 & 0.13 \\
2451 & 385 & 47000 & 7200 & 42900 & 1700 & 3.8 & 0.2 & 3.87 & 0.10 & 5.559 & 0.120 & 5.55 & 0.29 \\
1008 & 511 & 42000 & 1000 & 43700 & 1700 & 4.0 & 0.1 & 4.25 & 0.11 & 5.365 & 0.049 & 5.46 & 0.15 \\
1699 & 667 & 41000 & 1000 & 38750 & 820 & 3.9 & 0.1 & 3.59 & 0.10  & 5.284 & 0.050 & 5.21 & 0.10 \\
1433 & 599 & 40000 & 4500 & 47300 & 500 & 3.6 & 0.1 & 4.02 & 0.10  & 5.783 & 0.066 & 6.01 & 0.10 \\
1890 & 601 & 40000 & 1000 & 40280 & 500 & 3.8 & 0.1 & 3.94 & 0.10  & 5.495 & 0.046 & 5.55 & 0.18 \\
276 & 491 &  39000 & 1400 & 40360 & 800 & 3.6 & 0.2 & 3.84 & 0.10   & 5.415 & 0.049 & 5.43 & 0.16 \\
2780 & 648 & 39000 & 1000 & 40000 & 1500 & 3.6 & 0.1 & 3.80 & 0.10 & 5.607 & 0.049 & 5.66 & 0.13 \\
955 & 536 &  39000 & 1400 & 41500 & 1540 & 4.0 & 0.2 & 4.23 & 0.15  & 5.124 & 0.048 & 5.19 & 0.17 \\
195 & 494 &  37000 & 1200 & 38940 & 1740 & 3.8 & 0.2 & 4.21 & 0.21  & 4.993 & 0.046 & 5.03 & 0.20 \\
2385 & 456 & 36000 & 12400 & 35850 & 640 & 3.4 & 0.4 & 3.93 & 0.10 & 5.117 & 0.063 & 5.17 & 0.10 \\
3081 & 484 & 36000 & 1200 & 35680 & 680 & 3.4 & 0.1 & 3.68 & 0.10  & 5.276 & 0.046 & 5.41 & 0.14 \\
3200 & 564 & 36000 & 1800 & 37000 & 1500 & 4.1 & 0.4 & 4.10 & 0.10 & 4.818 & 0.047 & 5.33 & 0.13 \\
1870 & 611 & 35000 & 1700 & 37410 & 900 & 3.6 & 0.3 & 4.13 & 0.14  & 4.724 & 0.047 & 4.79 & 0.14 \\
2223 & 436 & 35000 & 1900 & 35000 & 1500 & 3.8 & 0.4 & 3.90 & 0.10 & 4.463 & 0.049 & 4.87 & 0.13 \\
2897 & 419 & 34000 & 1000 & 33100 & 900 & 3.6 & 0.2 & 3.64 & 0.10  & 5.040 & 0.046 & 5.07 & 0.24 \\
774 & 664 &  34000 & 1200 & 35700 & 500 & 3.2 & 0.1 & 3.58 & 0.10   & 5.500 & 0.170 & 5.53 & 0.10 \\
375 & 597 &  34000 & 1000 & 35400 & 720 & 3.6 & 0.1 & 3.94 & 0.11   & 4.826 & 0.047 & 4.87 & 0.14 \\
2985 & 571 & 33000 & 1300 & 31100 & 770 & 4.1 & 0.3 & 4.31 & 0.10  & 4.383 & 0.048 & 4.39 & 0.10 \\
1297 & 493 & 33000 & 2600 & 37050 & 950 & 3.3 & 0.5 & 4.27 & 0.10  & 4.112 & 0.061 & 5.06 & 0.16 \\
3172 & 609 & 33000 & 2200 & 33000 & 1500 & 3.9 & 0.5 & 3.82 & 0.10 & 4.401 & 0.047 & 4.52 & 0.13 \\
911 & 498 &  33000 & 1000 & 33230 & 810 & 3.9 & 0.2 & 4.12 & 0.15   & 4.813 & 0.049 & 4.88 & 0.14 \\
1334 & 635 & 32000 & 1600 & 34120 & 500 & 3.5 & 0.3 & 4.00 & 0.10  & 4.877 & 0.046 & 4.83 & 0.12 \\
1387 & 592 & 32000 & 1800 & 33560 & 1000 & 4.0 & 0.3 & 4.28 & 0.13 & 4.550 & 0.063 & 4.69 & 0.13 \\
2038 & 649 & 32000 & 1500 & 34750 & 630 & 3.6 & 0.2 & 4.19 & 0.10  & 4.607 & 0.047 & 4.71 & 0.12 \\
2256 & 393 & 32000 & 1200 & 31600 & 500 & 3.5 & 0.2 & 3.55 & 0.10  & 4.882 & 0.048 & 4.92 & 0.10 \\
3027 & 660 & 32000 & 1400 & 32260 & 1020 & 3.8 & 0.2 & 4.15 & 0.16 & 4.708 & 0.047 & 4.73 & 0.20 \\
710 & 560 &  31000 & 1900 & 33570 & 1150 & 3.6 & 0.4 & 4.20 & 0.16  & 4.475 & 0.047 & 4.52 & 0.18 \\
2815 & 620 & 31000 & 2700 & 31700 & 830 & 3.8 & 0.4 & 4.11 & 0.10  & 4.581 & 0.061 & 4.31 & 0.10 \\
2114 & 607 & 30000 & 1900 & 32800 & 560 & 3.6 & 0.3 & 4.23 & 0.10  & 4.437 & 0.059 & 4.56 & 0.10 \\
1826 & 624 & 30000 & 1500 & 29000 & 1080 & 3.8 & 0.3 & 4.00 & 0.44 & 4.348 & 0.061 & 4.20 & 0.10 \\
2939 & 554 & 30000 & 2900 & 34130 & 770 & 3.7 & 0.5 & 4.30 & 0.10  & 4.256 & 0.082 & 4.51 & 0.10 \\
1894 & 659 & 28000 & 1600 & 30000 & 1920 & 3.5 & 0.2 & 4.30 & 0.10 & 4.473 & 0.060 & 4.55 & 0.11 \\
888 & 612 &  28000 & 2200 & 27000 & 1000 & 3.7 & 0.2 & 4.30 & 0.10  & 4.481 & 0.077 & 4.48 & 0.10 \\
3018 & 449 & 28000 & 13500 & 24000 & 1000 & 3.8 & 1.2 & 3.80 & 0.10& 3.964 & 0.184 & 3.87 & 0.10 \\
2804 & 575 & 26000 & 2100 & 26000 & 1000 & 3.3 & 0.2 & 3.75 & 0.20 & 4.704 & 0.062 & -- & -- \\
1951 & 646 & 25000 & 1600 & 24000 & 1000 & 2.6 & 0.1 & 2.80 & 0.10 & 4.853 & 0.160 & 4.77 & 0.10 \\
1689 & 420 & 25000 & 2100 & 26500 & 1000 & 2.6 & 0.2 & 3.00 & 0.20 & 5.812 & 0.194 & 5.84 & 0.10 \\
2190 & 590 & 23000 & 1100 & 24000 & 1000 & 2.5 & 0.1 & 2.80 & 0.20 & 5.786 & 0.150 & 5.87 & 0.10 \\
1399 & 417 & 21000 & 4500 & 18500 & 1000 & 2.7 & 0.5 & 2.55 & 0.20  & 4.822 & 0.160 & 4.51 & 0.20 \\
                        \hline
                \end{tabular}
\tablefoot{Columns 1 and 2: MUSE and VFTS identifications (Paper I and \citealt{2011A&A...530A.108E}, respectively).
Columns 3-6: Effective temperatures (and uncertainties) from MUSE
and published studies from the VFTS \citep[][]{2014A&A...564A..39S,2017A&A...601A..79S,2015A&A...575A..70M,2017A&A...600A..81R,2017A&A...603A..91G,2018Sci...359...69S}.
Columns 7-10: Same, but for surface gravities. Columns 11-14: Same, but for luminosities.}
\end{table*}

\subsection{Spectral classification of MUSE spectra}
\label{Sect:SpT_cal}

We used a sample of
stars available in both the MUSE data and the VFTS to investigate the use of the
\ion{He}{i}$\lambda4921$ and \ion{He}{ii}$\lambda5411$ lines for
spectral classification.  If these are seen to compare well with the
classical criteria at shorter wavelengths, they could enable
the classification of massive stars observed in other MUSE programmes \cite[see also][]{2020A&A...634A..51B}.

\begin{figure}[]
        \resizebox{\hsize}{!}{\includegraphics[angle=0,width=\textwidth]{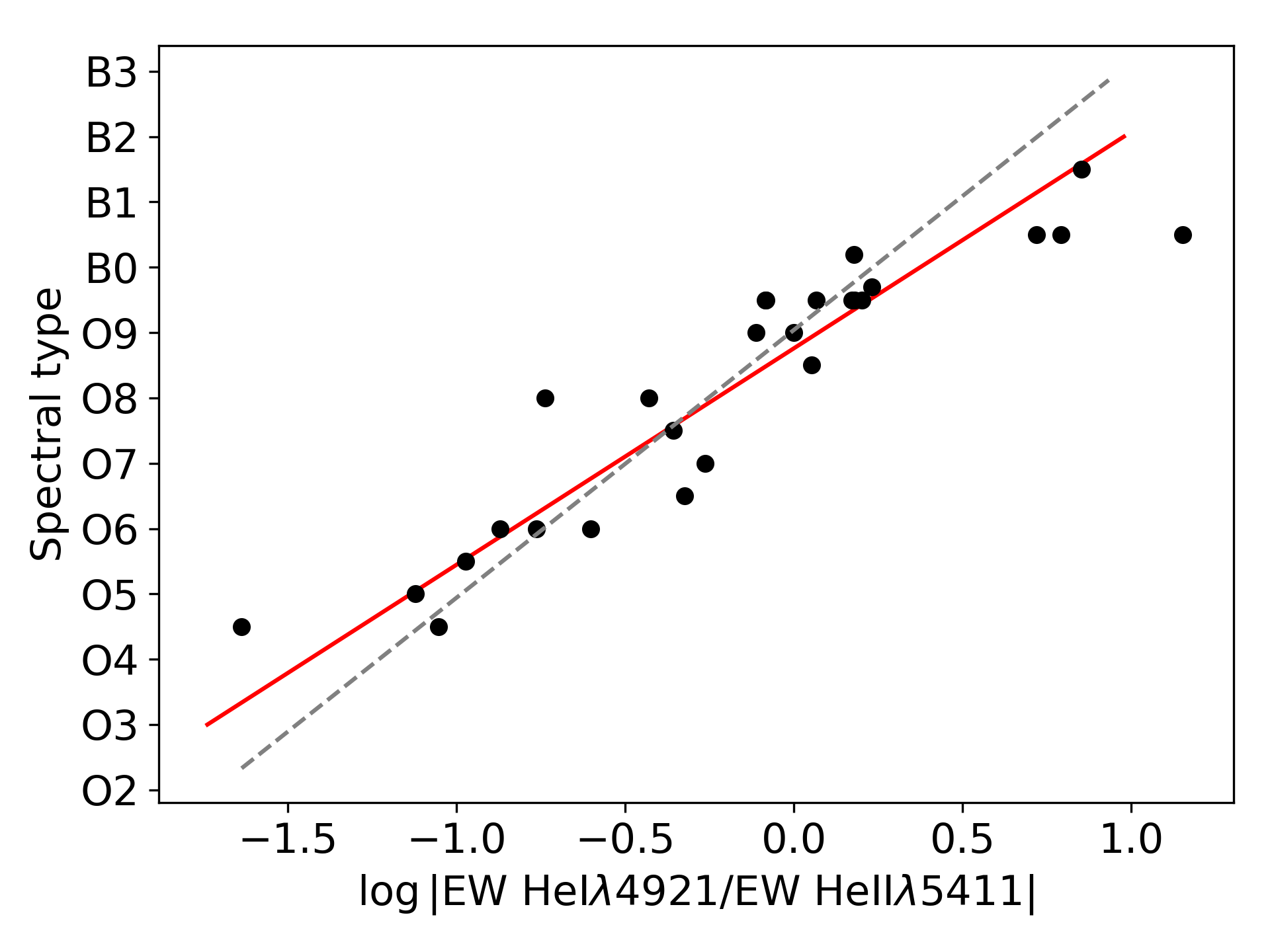}}
        \caption{Spectral type calibration (solid red line) from
          fitting the \ion{He}{i}$\lambda 4921$/\ion{He}{ii}$\lambda 5411$ equivalent-width ratio
          of the stars observed with MUSE that were classified from
           VFTS data \citep{2014A&A...564A..40W,2015A&A...574A..13E}.
           The calibration from
          \cite{1999AJ....117.2485K}, based on the same \ion{He}{i}$\lambda 4921$/\ion{He}{ii}$\lambda 5411$ ratio for Galactic O stars, is also shown (dashed grey line).}\label{Fig:SpT_Cal}
\end{figure}

Given the resolution of the MUSE data, we approached these
measured equivalent widths (rather than line-intensity ratios)
by comparing them to the spectral classifications from the VFTS
\citep{2014A&A...564A..40W,2015A&A...574A..13E}. The relationship
between equivalent width and spectral type is luminosity-class-dependent.
The sample is mainly populated by luminosity class
V, and only these stars were used in the calibration.  A relatively tight
relationship was found, as shown in Fig.~\ref{Fig:SpT_Cal}, with a
linear fit given by:
 SpT\,$=$\,3.31\,[log(\ion{He}{i}/\ion{He}{ii})]\,$+$\,8.76\footnote{SpT:
  O2\,$=$\,2...O9\,$=$\,9, B0\,$=$\,10...B9\,$=$\,19}.  The typical
uncertainty for a given equivalent-width ratio is of the order of
$\pm$1 subtype, which is usually sufficient for the sort of
exploratory studies undertaken with MUSE. We find a similar trend to
that reported by \cite{1999AJ....117.2485K} for Galactic O stars.  In particular, their
results over the range
$-$1\,$<$\,[log($\ion{He}{i}$/$\ion{He}{ii}$)]\,$<$\,1 are a reasonable
match to our LMC stars.

\begin{figure}[]
        \resizebox{\hsize}{!}{\includegraphics[angle=0,width=\textwidth]{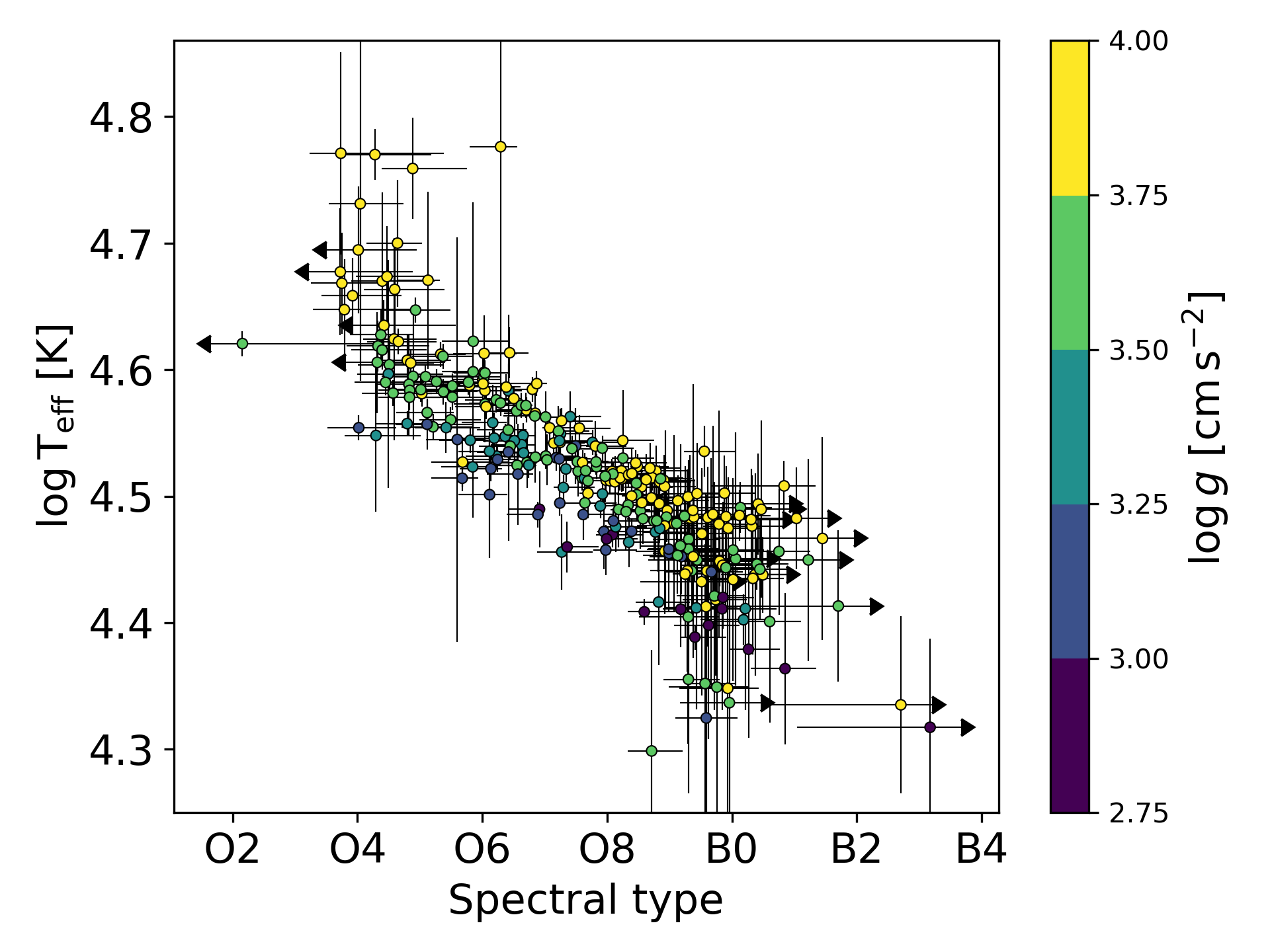}}
        \caption{Estimated effective temperatures (T$_{\rm eff}$) of the
          MUSE stars versus their spectral types from the calibration in
          Fig.~\ref{Fig:SpT_Cal}. For clarity, the stars have been
          artificially distributed into (continuous) spectral types to
          avoid significant overlap. The distribution is colour-coded
          by surface gravity (see
          Table~\ref{TAB:cat}).}\label{Fig:SpT_Teff}
\end{figure}

We measured the \ion{He}{i}/\ion{He}{ii} ratio of the full MUSE
sample and used these to estimate spectral types, as listed in
Table~\ref{TAB:cat}.  Figure~\ref{Fig:SpT_Teff}
compares the T$_{\rm eff}$ estimates with the classifications obtained
from the \ion{He}{i}/\ion{He}{ii} ratio (again excluding those with
\teff\,$<$\,4.3; see Sect.~\ref{Sect:post}). The trend between the
estimated surface gravities and spectral types suggests a
dependence on luminosity class \citep[see][]{2014A&A...570L...6S}.
However, our calibration \citep[and that from][]{1999AJ....117.2485K}
is based on dwarf (luminosity class V) stars, so we are unable
to investigate luminosity effects at this point (see
Fig.~\ref{Fig:SpT_Cal}). Not unexpectedly, the largest discrepancies
with the linear trend are found at the extremes of the distribution where the diagnostic lines become very weak or absent (i.e.
\ion{He}{ii}$\lambda$4921 at the earliest O-types, and
\ion{He}{ii}$\lambda$5411 for B0 and later).

\section{Evolutionary status of NGC\,2070}
\label{Sect:shrd}

\begin{figure*}[]
        \resizebox{\hsize}{!}{\includegraphics[angle=0,width=\textwidth]{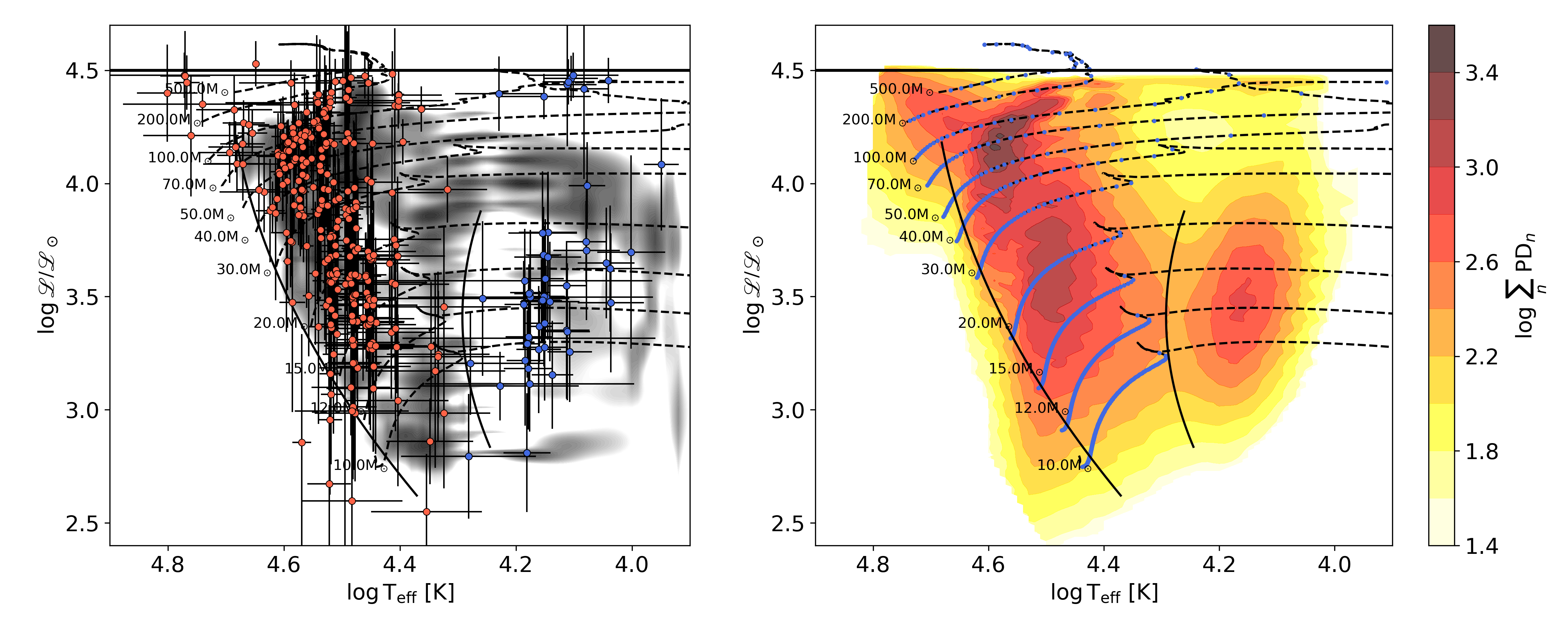}}
        \caption{Spectroscopic HRD (sHRD, ${\mathscr L}$\,$\equiv$\,T$_{\rm
        			eff}^4/g$, \citealt{2014A&A...564A..52L}) for the MUSE sample.
       Left: Positions of the MUSE sample in the sHRD (red and blue dots, offset slightly where
       required to avoid overlap) compared to the equivalent
          probability distribution (PD) for the Milky Way from
          \citet[][grey shaded regions]{2014A&A...570L..13C} and
          theoretical rotating evolutionary tracks from
          \citet{2015A&A...573A..71K}. Stars apparently beyond the TAMS, flagged as candidate
          emission-line stars \citep{2018ApJ...868...57C}, are marked in blue in the left-hand panel (see
          Sect.~\ref{Sect:post}).  {\it Right:} Summed
          PD functions from the MUSE analysis
          and the same evolutionary tracks, in which the blue dots
          indicate equal time steps of 0.1\,Myr. Horizontal lines
          at log\,\lpr\,$=$\,4.5 mark the upper limit of the model
          grid (the Eddington limit is at log\,\lpr\,$=$\,4.6). The
          other two solid black lines indicate the empirical ZAMS and
          TAMS from \cite{2014A&A...570L..13C} for the Milky Way.}\label{Fig:SAMPLE}
\end{figure*}

The sHRD for the MUSE sample is shown in Fig.~\ref{Fig:SAMPLE} and can
be described by two groups.  The largest fraction (281 of 333 stars)
is located in the region of the main sequence predicted by the
evolutionary models, accounting for the effects of rotation from
\cite{2015A&A...573A..71K}. There are two distinctive populations in
the main sequence, as highlighted by the right-hand panel of
Fig.~\ref{Fig:SAMPLE}, which shows the summed probability distributions
from the analysis of each star along the entire {\sc fastwind} grid \citep[see][]{2018ApJ...868...57C}.
The remaining 52 stars appear to lie beyond  the theoretical TAMS at \teff\,$<$\,4.3 from \cite{2015A&A...573A..71K}.

\begin{figure*}[]
        \resizebox{\hsize}{!}{\includegraphics[angle=0,width=\textwidth]{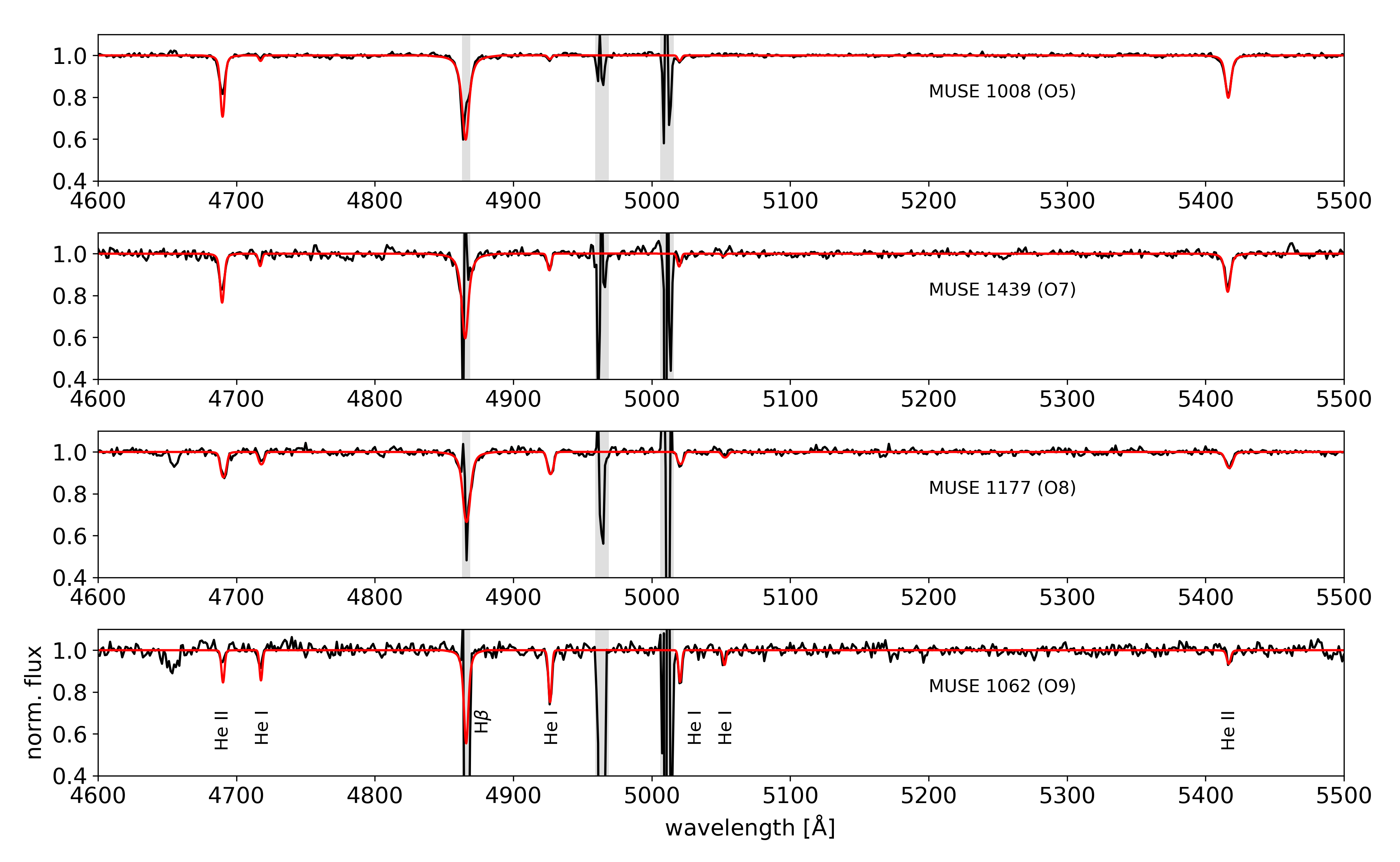}}
        \caption{Examples of model fits to four O-type main sequence stars. Areas
                strongly affected by nebular contamination are marked in grey.
Spectral types based on \ion{He}{i}$\lambda4921$ and \ion{He}{ii}$\lambda5411$
equivalent widths are also included (see Sect.~\ref{Sect:SpT_cal}). As expected,
the  \ion{He}{i}$\lambda4921$ line increases as we move towards later spectral types,
while the \ion{He}{ii}$\lambda5411$ decreases, thus supporting our algorithm for
spectral classification. }\label{Fig:GROUP_1}

\bigskip

        \resizebox{\hsize}{!}{\includegraphics[angle=0,width=\textwidth]{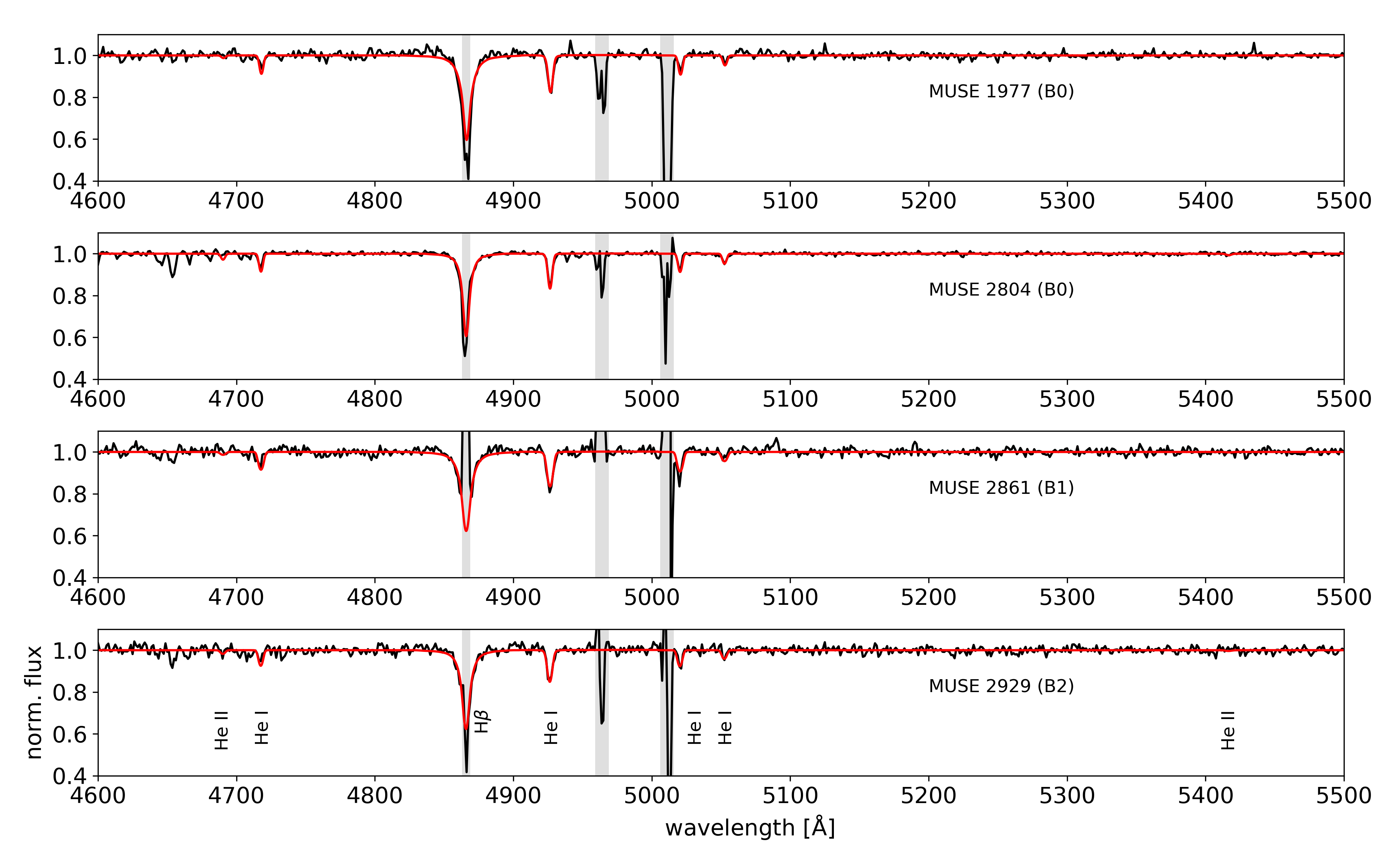}}
        \caption{Example of model fits to four early B-type main sequence stars (see Fig.~\ref{Fig:GROUP_1}).}\label{Fig:GROUP_2}
\end{figure*}

\subsection{Main-sequence stars}
\label{Sect:main-sequence}

Figure~\ref{Fig:SAMPLE} suggests two subgroups within the
main sequence sample, which may reflect  different bursts of
star-formation in the region \citep[e.g.][and also
Sect.~\ref{Sect:Age}]{2015ApJ...811...76C}. \ion{He}{ii}\,$\lambda$5411  is clearly seen in 147 stars (see
Fig.~\ref{Fig:GROUP_1}), with the distribution of estimated
temperatures peaking at \teff$\,\sim$4.55. Only a few stars are found
with \teff\,$>$\,4.6, closer to the theoretical ZAMS.  The distribution of
these very hot stars matches the expected position of merger
candidates \citep{2016MNRAS.457.2355S} and/or binary evolution
products \citep{2020ApJ...888L..12W}. However, these stars also have large uncertainties 
in their parameters as they are constrained only by \ion{He}{ii} and H$\beta$ and 
the non-detection of \ion{He}{i} lines.

Given that star formation is still underway in NGC\,2070
\citep{1999AJ....117..225W,2013AJ....145...98W},
 we would have expected to find a larger population close to the
expected ZAMS.  The lack of massive
($>$\,30\,M$_\odot$) O-type stars close to the ZAMS has previously been noted in the Milky Way
\citep{2014A&A...570L..13C,2018A&A...613A..65H,2020A&A...638A.157H} and in the SMC
\citep{2013ApJ...763..101L,2018ApJ...868...57C}. It was suggested that
very young stars may still be embedded in their natal clouds and thus not
accessible for optical observations. However, \cite{1984ApJ...287..116K} has
shown that the embedded phase is expected to be relatively
short (10\% of the typical $\sim$5\,Myr lifetime).
\cite{1986ARA&A..24...49Y} claimed that the pre-main sequence
contraction timescale becomes shorter than the accretion timescale
for the highest-mass stars; in this scenario, the most massive stars
would ignite hydrogen while still accreting.
Moreover, the current sample does not include  \ion{He}{ii}$\lambda5411$ emission stars, and the R136
stellar cluster is not resolved and so is missing from this analysis.
R136 cluster stars \citep{2020MNRAS.499.1918B} and \ion{He}{ii}$\lambda5411$
emission-line stars must be quantified before further conclusions can be reached
regarding the small number of the most massive stars near the ZAMS  \citep[see also][]{2016MNRAS.458..624C}.

Further down the main sequence, we find 134 late O-type and early B-type stars with
typical temperatures of \teff\,$\sim$4.5. Example spectra and their
model fits are shown in Fig.~\ref{Fig:GROUP_2}, in which the
\ion{He}{i} lines dominate the spectrum; weak
\ion{He}{ii}~$\lambda$4686 absorption in a couple of cases indicates that
they are on the cusp of the O-B transition at the very earliest B-types. These stars are still too young to have reached
the TAMS and cannot serve as reliable anchors such as those proposed
in the Milky Way \citep{2014A&A...570L..13C} and the SMC
\citep{2018ApJ...868...57C}.

The region of the sHRD in Fig.~\ref{Fig:SAMPLE} where we would expect
to find B-type supergiants (\teff$\sim4.1$, log\,\lpr\,$ > 4.0$)
includes several stars, matching  the extended main sequence
predicted by \cite{2015A&A...573A..71K}. The extension of the
empirical TAMS in the upper part of the sHRD suggests a possible
envelope inflation scenario
\cite[see][]{2015A&A...580A..20S,2017A&A...597A..71S}. However, given
the young age of NGC\,2070, the sample of presumed supergiants in the
MUSE data is too small to provide robust tests of these predictions.
We add that our results place these stars at the edge of the model
grid, and as such additional analysis is warranted.

\subsection{Stars apparently beyond the theoretical TAMS}
\label{Sect:post}

\begin{figure*}[]
        \resizebox{\hsize}{!}{\includegraphics[angle=0,width=\textwidth]{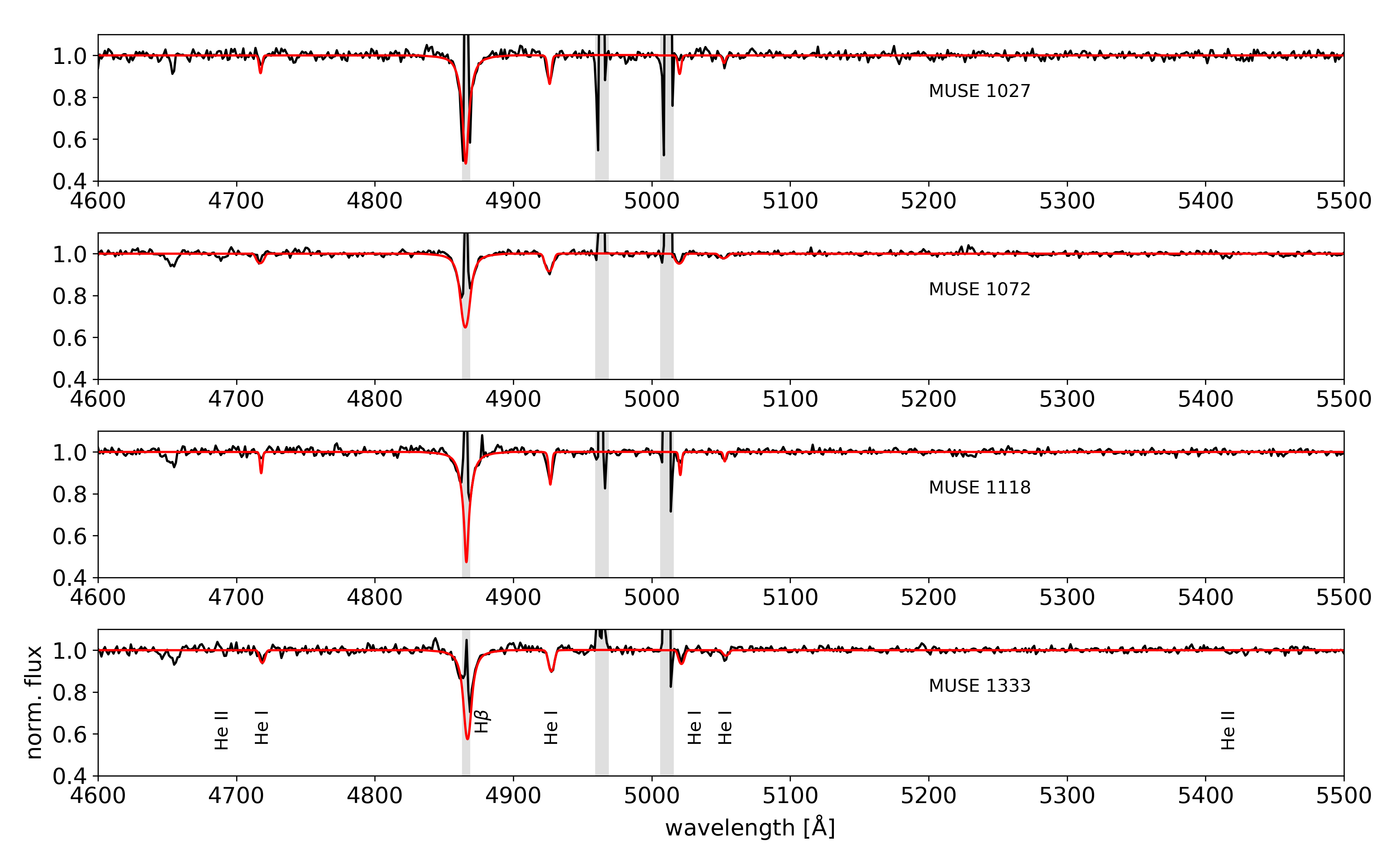}}
        \caption{Examples of model fits to four stars apparently
          beyond the TAMS from the model-atmosphere analysis but
          which we suspect are emission-line objects and/or
          contaminated by significant nebular emission (regions marked in grey),
          such that the   estimated temperatures are unreliable (see Sect.~\ref{Sect:post} and
           \citealt{2018ApJ...868...57C}).}\label{Fig:GROUP_3}
\end{figure*}

We find 52 stars (including the B-type supergiants) with
\teff\,$<4.3$, thus placing them beyond the TAMS predicted by the
models of \cite{2015A&A...573A..71K}.  The spectra of four examples of
this group are shown in Fig.~\ref{Fig:GROUP_3}. Once stars start
burning He in their cores, theoretical evolutionary tracks predict a
rapid evolution until they expand and cool to reach the red supergiant
(RSG) phase.  The position of these objects between the main sequence
and the RSG phase is therefore puzzling and not in agreement with the
expected evolution of a single star, unless they are blue-loop objects
in a post-RSG phase \citep[e.g.][]{2019A&A...625A.132S}.

The sHRD for the Milky Way from \cite{2014A&A...570L..13C} also found
stars in this range (\teff\,$\approx$4.1), with
\cite{2018ApJ...868...57C} finding that most of these objects were
emission-line stars (see Fig. 8 in \citealt{2018ApJ...868...57C}).
The potential contribution from circumstellar material is not taken into account in the atmospheric analysis, which may be influencing the
results \citep[e.g. via dilution;][]{2015A&A...578A..26C}.

The examples in Fig.~\ref{Fig:GROUP_3} display emission in H$\beta$,
although [\ion{O}{iii}] emission is also seen, suggesting that the
H$\beta$ emission is, at least in part, due to nebular contamination.
In contrast to the \cite{2018ApJ...868...57C} study, here we cannot
 disentangle nebular contamination from any other contribution
in the H$\beta$ and H$\alpha$ emission lines. We also note that two
objects in Fig.~\ref{Fig:GROUP_3} show weak
\ion{He}{ii}$\lambda$4686 absorption, which is indicative of substantially
higher temperatures than our estimates and confirms our suspicions
regarding the reliability of the temperatures (although there is no
corresponding \ion{He}{ii}$\lambda$5411 absorption in either). Therefore, these stars are not considered in the following discussion nor in Table~\ref{TAB:cat} and will
require further studies to shed light on their nature and properties.

\section{Discussion}\label{discussion}

\subsection{Age bi-modal distribution}
\label{Sect:Age}

\begin{figure*}[]
        \resizebox{\hsize}{!}{\includegraphics[angle=0,width=\textwidth]{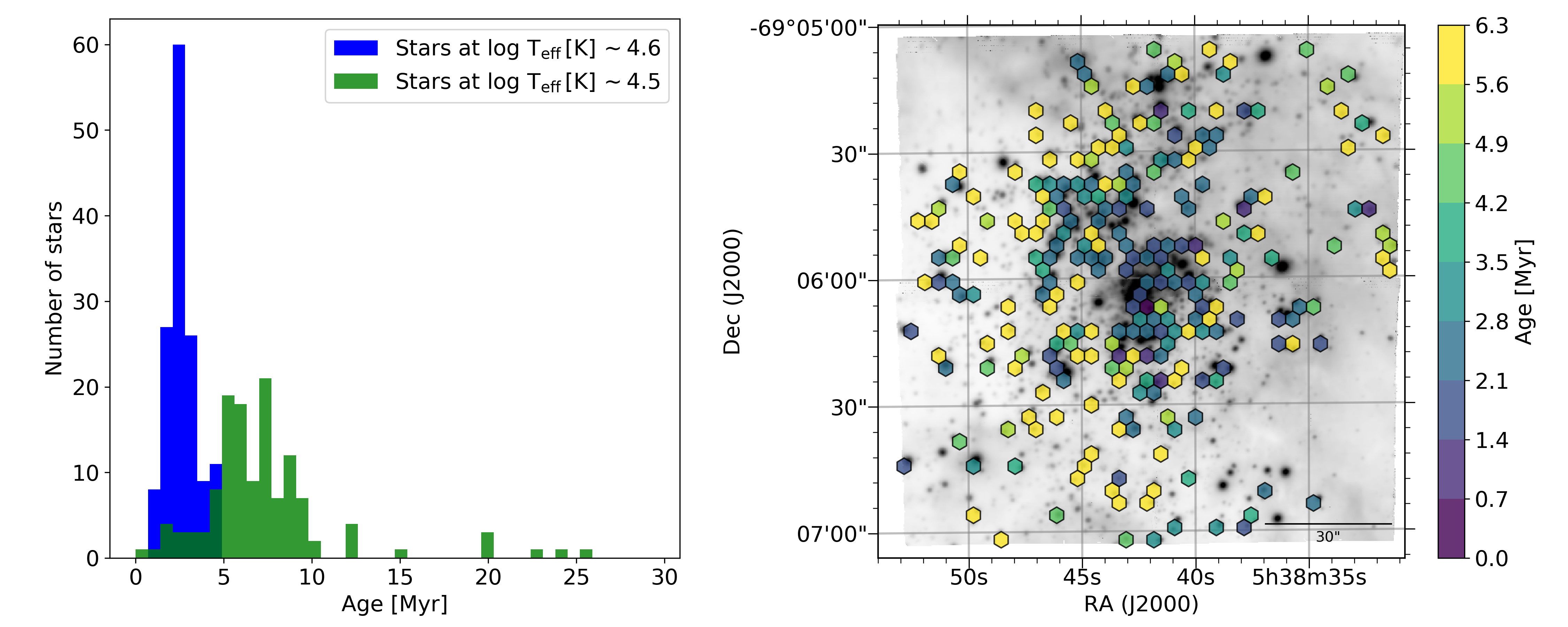}}
        \caption{	Distribution of inferred ages from the
          location of stars (\teff\,$>$\,4.3) in the sHRD and the
          evolutionary tracks from \cite{2015A&A...573A..71K} (left panel). The two
          main sequence groups highlighted in
          Sect.~\ref{Sect:main-sequence} are shown in different colours. {\it Right:} Spatial
          distribution of the sample, in which  each hexabin shows the
           mean age value of all stars within it.}\label{Fig:AGE}
\end{figure*}

The two well-populated parts of the main sequence in the sHRD
(Fig.~\ref{Fig:SAMPLE}) suggest two different stages of star formation
in the recent history of NGC\,2070.
Effective temperatures and gravities, resulting from the
spectral atmosphere analyses, were compared to the LMC evolutionary
tracks from \cite{2015A&A...573A..71K}.  Masses and ages were calculated by interpolating the tracks from
K\"ohler et al. at each target’s sHRD location as marked
by our results for effective temperatures and gravities, using SciPy
libraries\footnote{https://docs.scipy.org/doc/scipy/reference/interpolate.html}.
Ages are shown in the left-hand panel of Fig~\ref{Fig:AGE}; stars with
estimated parameters outside  the evolutionary tracks were not considered
further at this point.  We found an average age for the O-type
main sequence stars of 2.1\,$\pm$\,0.8\,Myr. The cooler group of main sequence
stars (centred on \teff\,$\sim$\,4.5) have an average age of 6.2\,$\pm$\,2\,Myr.
These results are in broad agreement with other age estimates for NGC\,2070
\citep{1997ApJS..112..457W,1998ApJ...509..879D,2012ApJ...754L..37S,2016MNRAS.458..624C,2018A&A...618A..73S}.
Gravity uncertainties may affect the ages extracted from the sHRD.  We tested
increasing the gravities by 0.3\,dex (see Sect.~\ref{sect:vfts}), finding a similar
distribution in agreement, within the errors, with the bi-modal age distribution
presented in  Fig~\ref{Fig:AGE}. As shown in the right-hand panel of Fig.~\ref{Fig:AGE},
the younger stars are more clustered around R136. However, there is no clear age segregation in NGC\,2070. Older ages
seem to be placed in the outskirts of the cluster; however, the number of analysed
stars in these regions is lower than in the core (Fig.~\ref{Fig:FoV}),
which may be the result of a statistical bias.

\subsection{Mass discrepancies between the sHRD and HRD}
\label{Sect:Masa}

\begin{figure*}[]
        \resizebox{\hsize}{!}{\includegraphics[angle=0,width=\textwidth]{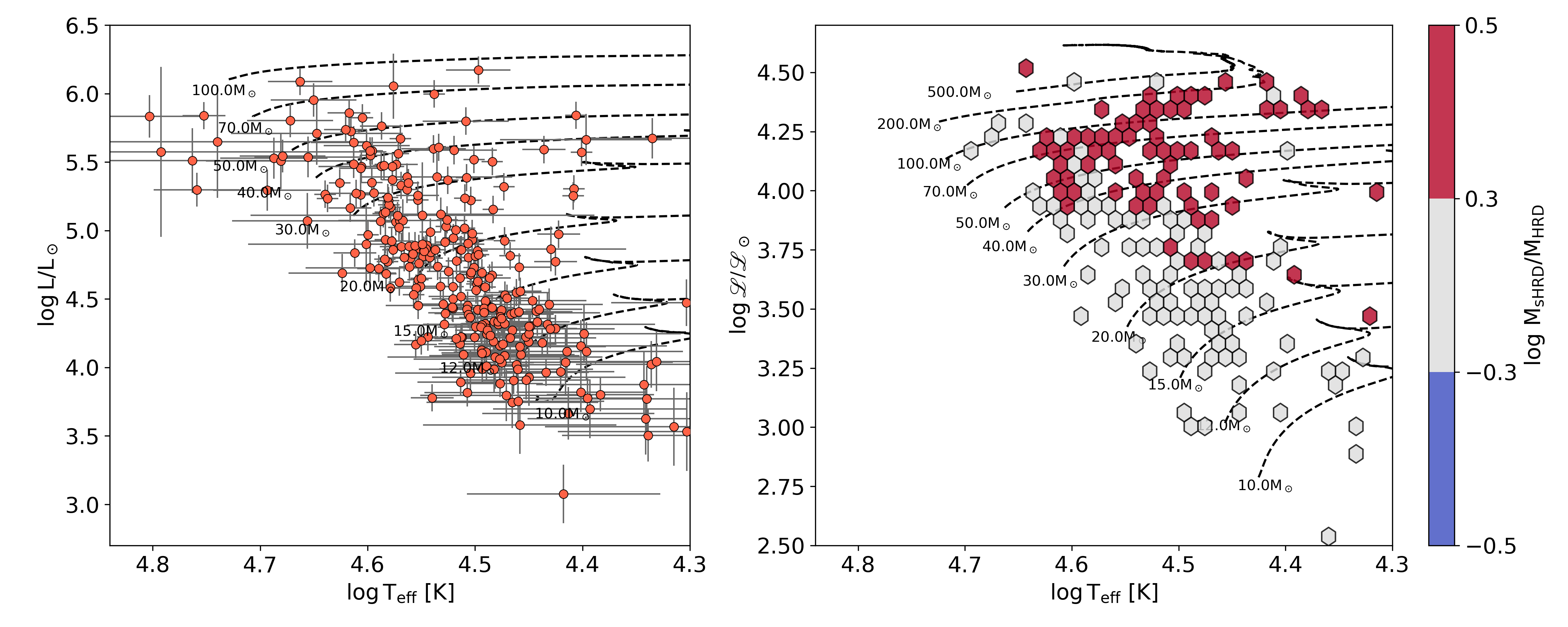}}
        \caption{
	        	 HRD for the MUSE sample with \teff\,$>$\,4.3 (left panel).
          {\it Right:} sHRD  (${\mathscr L}$\,$\equiv$\,T$_{\rm
                eff}^4/g$, \citealt{2014A&A...564A..52L}) for the sample indicating the difference
          in inferred mass between the sHRD and HRD approaches; each hexabin shows the mean value
                of all stars within it. Evolutionary tracks in both plots are those from
          \cite{2015A&A...573A..71K}. Stars outside of the evolutionary tracks are not
          included in the mass analysis.}\label{Fig:MASA}
\end{figure*}

The distribution of the MUSE sample in the HRD is shown in the
left-hand panel of Fig.~\ref{Fig:MASA}. Stellar luminosities were
calculated using the relevant photometry provided in Paper I (aside from the 13 stars marked with '*' in
Table~\ref{TAB:cat}) and adopting a distance to the LMC of 49.9\,kpc \citep{2013Natur.495...76P}.
Flux-calibrated MUSE spectra for each individual star were compared to the respective
synthetic  {\sc fastwind} spectral energy distributions (SEDs), obtained from the stellar
atmosphere analysis and parameters in Table~\ref{TAB:cat}. We estimated stellar radii and
colour excesses, $E(B-V)$, that provide the best match between the observed and synthetic
SEDs. We adopted an extinction law $Rv=Av/E(B-V)=4$ \citep{2014A&A...564A..63M}. We find a good
qualitative match between HRD and sHRD for the  late O-type and early B-type stars in the
main sequence, but there are differences for the most massive stars in the upper part of
the diagram. Those beyond the TAMS in the sHRD are systematically shifted to lower
luminosities and masses in the HRD. Due to the concern regarding the derived temperatures, stars
at \teff\,$<$\,4.3 are not included in the discussion pending further analyses.

We compared the masses inferred for our stars from their locations in
the sHRD and HRD. Masses were estimated according to their positions in both diagrams
and the \cite{2015A&A...573A..71K} LMC evolutionary tracks  using SciPy linear
interpolation libraries (see Sect.~\ref{Sect:Age}). Stars out of the parameter space
sampled by the evolutionary tracks were not included in this comparison.  At the high-mass end
($M$\,$\gtrsim$\,40\,$M_\odot$), the masses estimated from the sHRD
are significantly larger than the ones from the HRD,  as shown in the right-hand panel of
Fig.~\ref{Fig:MASA}.  These differences may be linked to the
long-standing discrepancy between spectroscopic and evolutionary mass
estimates for massive stars
\citep[e.g.][]{1992A&A...261..209H,2010A&A...524A..98W,2015IAUS..307..117M}.
A similar result was found from an analysis of the wider 30~Dor
population of O-type stars by \cite{2017A&A...601A..79S}. When using
the Kiel diagram (\teff\ and log\,$g$) to estimate masses
of similar stars in the Milky Way, they were also found to be larger than
those from the HRD at the high-mass end \citep{2018A&A...613A..12M}.  A similar
trend was found for the most massive stars in the SMC by \citet{2018ApJ...868...57C}.

The  uncertainty in the gravities from the MUSE data (Sect.~\ref{sect:vfts}) and
stellar wind constraints in the grid could have  contributed to the mass
differences: If the surface gravity is incorrect, simultaneously
fitting  T$_{\rm eff}$ and gravity can lead to underestimates of
T$_{\rm eff}$ \citep[e.g.][]{2017A&A...598A..60S}. The systematic difference of
approximately 0.3\,dex found for the stars present in both the VFTS
and MUSE data could somewhat alleviate this discrepancy.
That said, as similar mass discrepancies are seen in other studies (using different observations, algorithms, atmospheric models, etc.) it seems
unlikely that this is simply an artefact of the MUSE analysis.
We are now exploring alternative routes to improve the
estimates of stellar gravities from MUSE data, such as using absorption lines
from the hydrogen Paschen series (Bestenlehner et al. in prep.).

\subsection{Rotational velocities}
\label{Sect:Vsini}

\begin{figure*}[]
        \resizebox{\hsize}{!}{\includegraphics[angle=0,width=\textwidth]{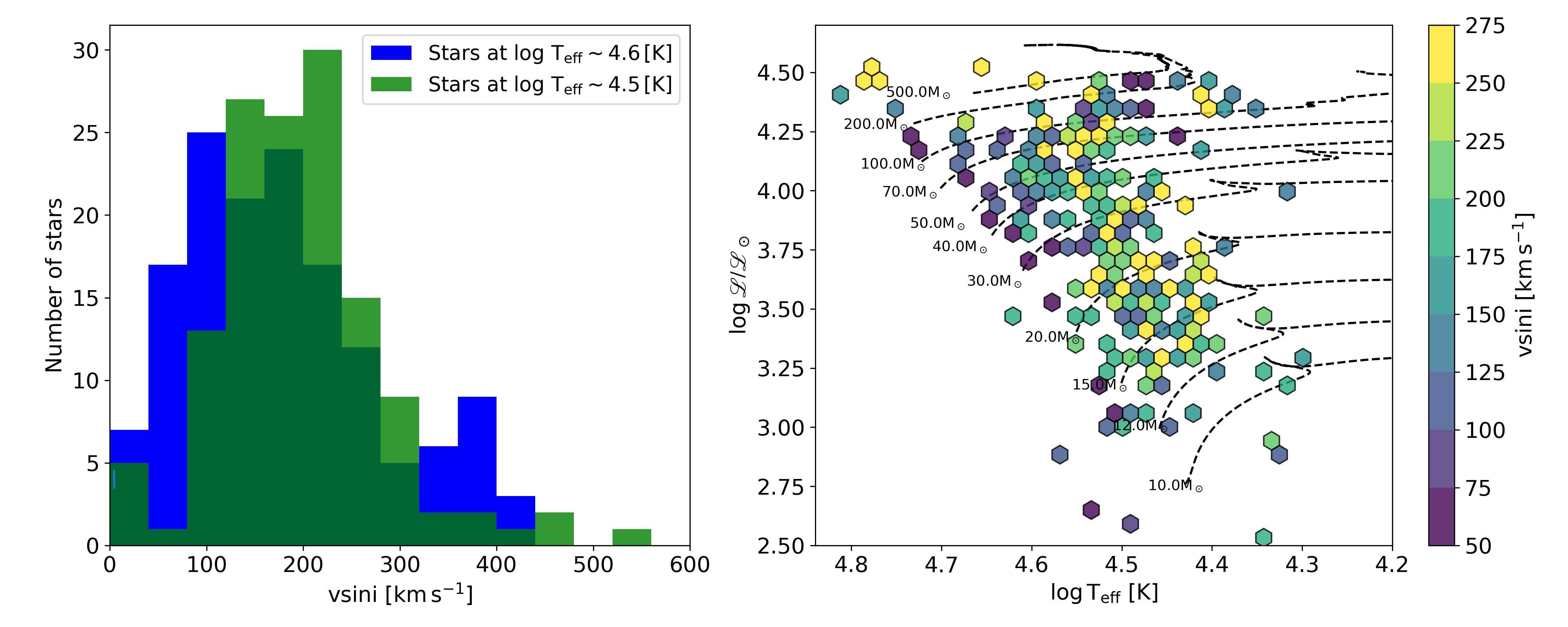}}
        \caption{ Distribution of projected rotational
          velocities (\vsini) for the two samples identified in the
          main sequence (left panel). {\it Right:} sHRD  (${\mathscr L}$\,$\equiv$\,T$_{\rm
                        eff}^4/g$) for the total sample, with
          \vsini\ average in each bin, and overlaid on the rotating evolutionary tracks from
          \cite{2015A&A...573A..71K}. Each hexabin shows the mean value
          of all stars within it.  }\label{Fig:vsini}
\end{figure*}

The distribution of estimated \vsini\ values is shown in the left-hand
panel of Fig.~\ref{Fig:vsini}. There is a peak at approximately
170\,\kms\ for the full sample, with no clear qualitative differences for the
stars separated into the two age groups  in the
main sequence. Their overall distribution resembles the  \cite{2013A&A...560A..29R}
results for the O-type stars in 30\,Dor rather than the \cite{2013A&A...550A.109D} bi-modal
distribution for lower-mass B-type stars.       The peak of the hot O-type star
group (\teff\,$\sim4.6$) also resembles the rotational velocity distribution found
in Cygnus\,OB2 in the Milky Way by \cite{2020A&A...642A.168B}. There is a relative
dearth of slow rotators in the cooler MUSE group, but we caution that the
spectral resolution of MUSE limits us to $\Delta$\vsini\,$\sim$\,60\,\kms\,,
and as such we are not able to robustly probe the low-velocity end of the distribution.
We note that \cite{2020MNRAS.492.2177K} have recently measured rotational velocities
for a total of 1400 stars of the intermediate age cluster NGC1846 in the LMC with uncertainties of typically 10 km/s using MUSE. A future, more refined
analysis may allow us to investigate these results in more detail.

There appears to be a prolongation in the distribution at
\vsini\,$\sim$\,370\,\kms\ for the hot O-type stars (see
Fig.~\ref{Fig:vsini}), which resembles the high-velocity tail found
for apparent single O-type stars in 30~Dor by
\citet{2013A&A...560A..29R}.  \cite{2013ApJ...764..166D} proposed that
rapidly rotating stars may originate from binary interactions and
mergers \citep{2014ApJ...782....7D,2016MNRAS.457.2355S}. These stars
in the MUSE sample merit further observations to test if these objects
are actually single stars as well as to investigate their physical properties
in more detail relative to the interaction and merger models. As shown in
the right-hand panel of Fig.~\ref{Fig:vsini}, there do not appear to
be strong trends in the \vsini\ estimates or their location in the
sHRD.

\section{Summary}
\label{Sect:Summ}

Exploiting the unique observational capabilities of MUSE, combined
with synthetic spectra calculated with {\sc fastwind}, we have
estimated physical parameters for 333 OB-type stars in NGC\,2070.
The majority of these objects are analysed for the first time here. Our
main conclusions can be summarised as follows.

\begin{itemize}
\item{281 stars (84\% of our sample) are still in the main sequence.
    They comprise a group of O-type stars with an average age
    of 2.1\,$\pm$\,0.8\,Myr and a group of late O-type and early
    B-type stars with an average age of 6.2\,$\pm$\,2.0\,Myr.  This
    spread in ages is consistent with previous studies, and we find no
    obvious age gradient across NGC\,2070.  }

\item{We find a relative dearth of O-type stars close to the
    theoretical ZAMS between approximately 20 and 50\,M$\odot$. Similar findings were reported  in the
    the Milky Way and the SMC  \citep{2018A&A...613A..65H,2020A&A...638A.157H,2018ApJ...868...57C}. Given the young
    age of NGC\,2070, this is somewhat unexpected, although stars in R136 are omitted from our sample.
    The stellar context of R136 needs to be better quantified before further
    conclusions can be reached \citep[see][]{2020MNRAS.499.1918B}.  Moreover, the stars close to the theoretical ZAMS,
    in the upper part of the sHRD ($>50$\,M$\odot$), match the predicted         position of stellar mergers \citep{2016MNRAS.457.2355S} and/or binary
        evolution products \citep[e.g.][]{2020ApJ...888L..12W}.
        However, the temperatures for these     stars have large uncertainties, and further
        studies of their properties are also required.}

\item{We find 52 stars, the rest of the analysed sample,        with temperatures beyond
        the theoretical TAMS from \cite{2015A&A...573A..71K} at \teff\,$<$\,4.3.         In the SMC study by   \cite{2018ApJ...868...57C}, a similar group of stars nearly all
    displayed signatures of Balmer emission.  This sub-sample in
    NGC\,2070 could similarly be emission-line stars and/or nebular contamination, where the
    analysis tools give incorrect results given the limitations of the
    data and the atmospheric models, but we need further observations to
    test this suggestion.}

\item{The HRD and sHRD are in good qualitative agreement for the late O-type and
    early B-type stars in the main sequence, but there are differences for
    the more massive O-type stars. Masses estimated from the sHRD
    compared to the evolutionary tracks are larger than those inferred
    from the HRD, which becomes severe (a factor of two to three) at
    $M$\,$>$\,40\,$M_\odot$.  This may be related to the uncertainty in
    the surface gravities estimated from the MUSE data
    (Sect~\ref{sect:vfts}). However, \cite{2017A&A...601A..79S} found
    similar problems at $M$\,$>$\,70\,$M_\odot$ in 30~Dor, and a
    comparable trend was also found by \citet{2018ApJ...868...57C} in
    the SMC (from observations of the more classical blue-visible
    range, albeit still at $R$\,$\sim$\,3000).}

\item{The projected \vsini\ distribution for the MUSE sample peaks at
    170\,\kms. We find a group of rapidly rotating O-type stars (with
    300\,$<$\,\vsini\,$<$\,450\,\kms) that resembles the high-velocity
    tail found for the wider population of O stars in 30~Dor by
    \citet{2013A&A...560A..29R}. Comparisons of our results with the
    bi-modal distribution for the B-type stars from
    \citet{2013A&A...550A.109D} are unfortunately limited by the
    velocity resolution of the MUSE data.}

\item{We used the \ion{He}{i}$\lambda4921$/\ion{He}{ii}$\lambda5411$
    equivalent-width ratio to estimate spectral types for our sample
    stars, calibrated using a subset of stars that overlap with the
    VFTS. In general, the effective temperatures estimated from the
    model-atmosphere fits follow a clear trend of decreasing
    temperature towards late spectral types.}
\end{itemize}

The next step beyond the current sample is analysis of the W-R and other \ion{He}{ii} emission-line stars
in NGC\,2070 to complete the distribution in the sHRD. Additional analyses to improve the quoted surface
gravities and overcome the limitations of basing the stellar gravity on H$\beta$ are  also desirable.
We are exploring the Paschen  lines as possible additional constraints for the
gravity (Bestenlehner in prep.). Furthermore, as with the rapidly rotating O-type stars
identified by Ramírez-Agudelo et al. (2013), the comparable group
from the MUSE data (Fig. 12) are interesting in the context of possible post-interaction or merger products,
and multi-epoch spectroscopy of this subgroup would be particularly valuable. As shown by \cite{2019A&A...632A...3G},
multi-epoch MUSE observations are indeed uniquely capable of detecting spectroscopic binaries in star clusters.

Finally, we note that the commissioning of the MUSE Narrow-Field Mode with adaptive
optics support has added a new capability for integral field spectroscopy from the ground
with angular resolution close to that from the Hubble Space Telescope. We have already secured
data for NGC2070 in this mode and will soon report on results in the most crowded regions
of R136 \citep{2021arXiv210201113C}.

In conclusion, integral field spectroscopy with MUSE has been demonstrated to be a powerful
tool for the quantitative spectroscopy of stars in crowded fields. Studies of this kind have only
begun to scratch the surface of what is expected to become an indispensable tool, in particular
in view of the upcoming generation of extremely large telescopes and the next generation of integral field
spectrographs (e.g. the BlueMUSE concept; \citealt{2019arXiv190601657R}).



\begin{acknowledgements}
The authors thank the referee for useful comments and helpful suggestions that improved
this manuscript. NC gratefully acknowledge funding from the
Deutsche Forschungsgemeinschaft (DFG) - CA 2551/1-1. SS-D and AHD acknowledge support from
the Spanish Government Ministerio de Ciencia e Innovaci\'on through
grants PGC-2018-091\,3741-B-C22 and CEX2019-000920-S, and from the Canarian Agency
for Research, Innovation and Information Society (ACIISI), of the Canary Islands
Government, and the European Regional Development Fund (ERDF), under grant with
reference ProID2020010016. Our research used Astropy, a community-developed core Python
package for Astronomy \citep{2013A&A...558A..33A}, and APLpy, an
open-source plotting package for Python \citep{2012ascl.soft08017R}.
\end{acknowledgements}



\bibliographystyle{aa}
\bibliography{AA_2020_40008}


\clearpage
\onecolumn

        \begin{landscape}

                \begin{longtable}{rrrrrrrrrrrrrrrr}
                        \caption{\label{TAB:cat} Estimated physical parameters of the 281 early-type stars with \teff\,$>$\,4.3 in NGC\,2070 from the MUSE observations sorted by effective temperature.  See Sect.\ref{Sect:Analisis} for a description of the analysis and limitations in the listed parameters. Columns 1-3: Identification and coordinates from \cite{2018arXiv180201597C}. Columns 4-9: Estimated effective temperatures, spectroscopic luminosities (${\mathscr L}$\,$\equiv$\,T$_{\rm eff}^4/g$, \citealt{2014A&A...564A..52L}), luminosities ($L$), and their uncertainties. The probability distributions for stars in the upper part of the sHRD, close to the edge of the grid (approximately log\,${\mathscr L}/{\mathscr L}_\odot>4.3$), do not provide enough information to properly estimate uncertainties, and the values cited in this table must be considered as rough estimations. Columns 10-11: Extinction and visual magnitudes. Thirteen stars included in this work were not listed in \cite{2018arXiv180201597C}. These stars are marked by a '*' after their magnitudes. Columns 12-13: Masses estimated from the sHRD and HRD, respectively. Column 14: Ages estimated from positions in the sHRD. Column 15: Estimates of projected rotational velocities (\vsini) from the spectral fitting. Column 16: Spectral types derived according to the \ion{He}{i}$\lambda4921$ to \ion{He}{ii}$\lambda5411$ ratio.  Masses and ages are estimated from comparisons with evolutionary models from \cite{2015A&A...573A..71K}. Stars outside of the evolutionary track boundaries are not considered in the analysis ('--').}\\
                        \hline
                        \hline
                        MUSE ID & $\alpha$ (J2000) & $\delta$ (J2000) & log\,$T_{\rm eff}$ & $\sigma_{log\,T_{\rm eff}}$ & log\,${\mathscr L}$ & $\sigma_{log\,{\mathscr L}}$ & log\,$L$ & $\sigma_{log\,L}$ & Av & V & Mass$_{\mathscr L}$ & Mass$_L$ & Age$_{\mathscr L}$ & $v \sin i$ & SpT \\
                        & [${\rm h,m,s}$] & [${\circ,',''}]$ & [K] & [K] & [${\mathscr L}_\odot$] & [${\mathscr L}_\odot$] & [$L_\odot$] & [$L_\odot$] & [mag] & [mag] & [$M_{\odot}$] & [$M_{\odot}$] & [Myr] & [km\,s$^{-1}$] & \\
                        \hline
                        \endfirsthead
                        \caption{Continued.} \\
                        \hline
                        MUSE ID & $\alpha$ (J2000) & $\delta$ (J2000) & log\,$T_{\rm eff}$ & $\sigma_{log\,T_{\rm eff}}$ & log\,${\mathscr L}$ & $\sigma_{log\,{\mathscr L}}$ & log\,$L$ & $\sigma_{log\,L}$ & Av & V & Mass$_{\mathscr L}$ & Mass$_L$ & Age$_{\mathscr L}$ & $v \sin i$ & SpT \\
                        & [${\rm h,m,s}$] & [${\circ,',''}]$ & [K] & [K] & [${\mathscr L}_\odot$] & [${\mathscr L}_\odot$] & [$L_\odot$] & [$L_\odot$] & [mag] & [mag] & [$M_{\odot}$] & [$M_{\odot}$] & [Myr] & [km\,s$^{-1}$] & \\
                        \hline
                        \endhead
                        \hline
                        \endfoot
                        \hline
                        \endlastfoot
                        1920 & 5:38:40.992 & -69:05:55.32 & 4.80 & 0.05 & 4.39 & 0.22 & 6.00 & 0.15 & 1.8 & 15.09 & -- & -- & -- & 150 & O4 \\
                        1537 & 5:38:39.864 & -69:06:07.92 & 4.77 & 0.21 & 4.48 & 0.20 & 5.74 & 0.63 & 1.8 & 15.51 & -- & -- & -- & 340 & O6 \\
                        830 & 5:38:39.288 & -69:06:39.24 & 4.77 & 0.04 & 4.48 & 0.20 & 6.02 & 0.13 & 3.2 & 16.20 & -- & -- & -- & 360 & O5 \\
                        2893 & 5:38:42.672 & -69:05:38.76 & 4.76 & 0.08 & 4.22 & 0.27 & 5.60 & 0.23 & 1.6 & 15.58 & -- & -- & -- & 50 & O4 \\
                        2245 & 5:38:39.48 & -69:05:10.32 & 4.76 & 0.02 & 4.45 & 0.12 & 5.76 & 0.07 & 1.2 & 14.78 & -- & -- & -- & 390 & O4 \\
                        1068 & 5:38:41.952 & -69:06:29.88 & 4.74 & 0.14 & 4.35 & 0.05 & 5.99 & 0.40 & 2.2 & 15.07 & -- & -- & -- & 130 & O4 \\
                        1401 & 5:38:42.048 & -69:06:16.92 & 4.69 & 0.05 & 4.15 & 0.11 & 5.40 & 0.15 & 1.6 & 15.58 & 70 & -- & 1.14 & 40 & O5 \\
                        2077 & 5:38:40.2 & -69:05:51.36 & 4.69 & 0.05 & 4.15 & 0.13 & 5.87 & 0.15 & 2.2 & 15.00 & 70 & 68 & 1.14 & 110 & O4 \\
                        2981 & 5:38:38.088 & -69:05:43.44 & 4.68 & 0.05 & 4.12 & 0.12 & 5.80 & 0.15 & 2.0 & 14.91 & 123 & 61 & 1.01 & 110 & O4 \\
                        2987 & 5:38:43.272 & -69:05:42.36 & 4.68 & 0.07 & 4.32 & 0.15 & 5.44 & 0.21 & 1.2 & 15.00 & 315 & -- & 0.46 & 210 & O4 \\
                        2901 & 5:38:32.304 & -69:05:44.52 & 4.67 & 0.07 & 4.08 & 0.16 & 5.33 & 0.20 & 2.0 & 16.01 & 104 & -- & 0.82 & 50 & O5 \\
                        1423 & 5:38:43.32 & -69:06:16.92 & 4.67 & 0.04 & 4.18 & 0.10 & 5.97 & 0.12 & 1.4 & 13.81 & 62 & 80 & 1.32 & 110 & O4 \\
                        2451 & 5:38:32.328 & -69:05:24 & 4.67 & 0.04 & 4.28 & 0.09 & 5.56 & 0.12 & 1.4 & 14.84 & 66 & 49 & 1.48 & 160 & O4 \\
                        2395 & 5:38:41.544 & -69:05:19.32 & 4.66 & 0.03 & 4.24 & 0.09 & 6.10 & 0.10 & 1.4 & 13.41 & 70 & 71 & 1.75 & 270 & O4 \\
                        1199 & 5:38:41.736 & -69:06:25.2 & 4.66 & 0.04 & 4.24 & 0.09 & 6.06 & 0.12 & 2.0 & 14.12 & 70 & 70 & 1.75 & 130 & O5 \\
                        1447 & 5:38:41.352 & -69:06:14.04 & 4.65 & 0.01 & 4.51 & 0.09 & 5.63 & 0.05 & 2.2 & 15.33 & 106 & 48 & 1.66 & 260 & O5 \\
                        1956 & 5:38:41.64 & -69:06:00.72 & 4.64 & 0.08 & 3.97 & 0.18 & 5.81 & 0.23 & 1.6 & 14.21 & 50 & 58 & 1.62 & 100 & O5 \\
                        1912 & 5:38:41.592 & -69:05:56.76 & 4.64 & 0.04 & 3.97 & 0.16 & 5.74 & 0.12 & 1.6 & 14.38 & 50 & 55 & 1.62 & 80 & O4 \\
                        2044 & 5:38:42.744 & -69:05:52.8 & 4.62 & 0.02 & 3.89 & 0.23 & 5.02 & 0.07 & 1.8 & 16.25 & 40 & -- & 2.19 & 20 & O4 \\
                        1008 & 5:38:41.736 & -69:06:28.08 & 4.62 & 0.01 & 3.89 & 0.11 & 5.36 & 0.05 & 1.6 & 15.18 & 40 & 35 & 2.19 & 110 & O5 \\
                        2760 & 5:38:45.696 & -69:05:38.76 & 4.61 & 0.01 & 4.14 & 0.07 & 5.74 & 0.05 & 1.4 & 13.97 & 146 & 50 & 1.19 & 90 & O4 \\
                        1570 & 5:38:42.432 & -69:06:06.48 & 4.61 & 0.02 & 3.84 & 0.39 & 5.88 & 0.07 & 1.4 & 13.62* & 40 & 60 & 1.85 & 50 & O6 \\
                        1523 & 5:38:39.888 & -69:06:09.36 & 4.61 & 0.11 & 4.14 & 0.19 & 5.00 & 0.29 & 1.6 & 16.03 & 146 & 28 & 1.19 & 350 & O6 \\
                        1699 & 5:38:52.824 & -69:06:12.24 & 4.61 & 0.01 & 3.94 & 0.10 & 5.28 & 0.05 & 1.4 & 15.12 & 97 & 34 & 1.35 & 40 & O5 \\
                        1963 & 5:38:42.696 & -69:05:56.04 & 4.61 & 0.04 & 4.24 & 0.11 & 5.96 & 0.11 & 2.6 & 14.62 & 100 & 68 & 1.77 & 90 & O4 \\
                        1535 & 5:38:39.696 & -69:06:08.64 & 4.61 & 0.02 & 4.04 & 0.12 & 5.34 & 0.07 & 1.8 & 15.37 & 121 & 36 & 1.27 & 200 & O5 \\
                        2946 & 5:38:37.704 & -69:05:42 & 4.61 & 0.01 & 4.04 & 0.10 & 5.49 & 0.05 & 2.0 & 15.19 & 121 & 43 & 1.27 & 130 & O5 \\
                        2112 & 5:38:44.976 & -69:05:54.24 & 4.61 & 0.02 & 4.14 & 0.08 & 5.66 & 0.07 & 1.8 & 14.58 & 146 & 47 & 1.19 & 150 & O4 \\
                        2819 & 5:38:44.424 & -69:05:36.24 & 4.61 & 0.01 & 4.14 & 0.06 & 5.92 & 0.05 & 1.6 & 13.72 & 146 & 64 & 1.19 & 110 & O2 \\
                        2177 & 5:38:44.16 & -69:05:56.76 & 4.61 & 0.02 & 4.14 & 0.13 & 6.00 & 0.07 & 2.0 & 13.94 & 146 & 75 & 1.19 & 50 & O4 \\
                        1433 & 5:38:46.224 & -69:06:17.64 & 4.60 & 0.02 & 4.20 & 0.07 & 5.78 & 0.07 & 1.4 & 13.80 & 89 & 59 & 2.39 & 110 & O5 \\
                        538 & 5:38:34.608 & -69:06:52.92 & 4.60 & 0.01 & 4.10 & 0.14 & 5.32 & 0.05 & 2.0 & 15.55 & 157 & 33 & 2.03 & 110 & O5 \\
                        1763 & 5:38:41.088 & -69:06:01.8 & 4.60 & 0.03 & 3.80 & 0.44 & 5.38 & 0.08 & 1.6 & 15.02 & 32 & 34 & 2.72 & 180 & O6 \\
                        2565 & 5:38:40.704 & -69:05:31.2 & 4.60 & 0.01 & 4.00 & 0.16 & 5.15 & 0.05 & 1.8 & 15.78 & 107 & 29 & 2.40 & 110 & O6 \\
                        1929 & 5:38:43.752 & -69:05:55.68 & 4.60 & 0.05 & 4.10 & 0.27 & 5.11 & 0.14 & 2.4 & 16.48 & 157 & 29 & 2.03 & 20 & O5 \\
                        1890 & 5:38:46.32 & -69:05:59.28 & 4.60 & 0.01 & 4.00 & 0.09 & 5.50 & 0.05 & 1.6 & 14.72 & 107 & 39 & 2.40 & 130 & O5 \\
                        1793 & 5:38:42.144 & -69:06:00.36 & 4.59 & 0.01 & 4.16 & 0.11 & 5.20 & 0.05 & 1.4 & 15.21 & 67 & 31 & 2.55 & 200 & O5 \\
                        2780 & 5:38:50.4 & -69:05:38.4 & 4.59 & 0.01 & 4.16 & 0.07 & 5.61 & 0.05 & 1.4 & 14.18 & 67 & 41 & 2.55 & 110 & O5 \\
                        1700 & 5:38:35.616 & -69:06:06.48 & 4.59 & 0.01 & 4.06 & 0.11 & 5.66 & 0.05 & 1.8 & 14.46 & 57 & 43 & 2.67 & 240 & O6 \\
                        1503 & 5:38:42.264 & -69:06:12.24 & 4.59 & 0.01 & 3.76 & 0.21 & 5.24 & 0.05 & 1.6 & 15.31 & 32 & 32 & 2.11 & 50 & O7 \\
                        276 & 5:38:40.848 & -69:06:57.6 & 4.59 & 0.01 & 4.16 & 0.11 & 5.42 & 0.05 & 2.2 & 15.46 & 67 & 35 & 2.55 & 110 & O5 \\
                        1502 & 5:38:42.36 & -69:06:07.92 & 4.59 & 0.02 & 3.46 & 0.48 & 5.65 & 0.06 & 1.8 & 14.48* & 25 & 43 & 0.30 & 180 & O6 \\
                        1494 & 5:38:43.128 & -69:06:11.52 & 4.59 & 0.01 & 4.16 & 0.08 & 5.66 & 0.05 & 1.6 & 14.26 & 67 & 43 & 2.55 & 170 & O6 \\
                        1781 & 5:38:42.864 & -69:05:59.28 & 4.59 & 0.06 & 4.36 & 0.14 & 5.06 & 0.16 & 2.2 & 16.34 & 91 & 28 & 2.27 & 160 & O6 \\
                        2003 & 5:38:40.32 & -69:06:00 & 4.59 & 0.09 & 4.46 & 0.05 & 6.12 & 0.24 & 1.6 & 13.08 & 106 & 82 & 1.96 & 270 & O4 \\
                        955 & 5:38:42.84 & -69:06:32.76 & 4.59 & 0.01 & 3.76 & 0.15 & 5.12 & 0.05 & 2.2 & 16.18 & 32 & 30 & 2.11 & 40 & O6 \\
                        2748 & 5:38:42.648 & -69:05:36.6 & 4.59 & 0.01 & 4.06 & 0.10 & 5.58 & 0.05 & 1.6 & 14.45 & 57 & 40 & 2.67 & 150 & O5 \\
                        1459 & 5:38:41.904 & -69:06:12.6 & 4.59 & 0.01 & 4.16 & 0.07 & 5.75 & 0.05 & 2.0 & 14.42 & 67 & 55 & 2.55 & 110 & O4 \\
                        3034 & 5:38:45.528 & -69:05:43.8 & 4.59 & 0.01 & 3.66 & 0.13 & 5.25 & 0.05 & 1.4 & 15.07 & 29 & 32 & 1.60 & 50 & O6 \\
                        1749 & 5:38:39.384 & -69:06:06.48 & 4.59 & 0.04 & 4.26 & 0.10 & 5.72 & 0.12 & 1.6 & 14.09 & 79 & 47 & 2.41 & 160 & O5 \\
                        963 & 5:38:42.696 & -69:06:36 & 4.59 & 0.01 & 4.16 & 0.08 & 5.62 & 0.05 & 2.0 & 14.75 & 67 & 42 & 2.55 & 220 & O5 \\
                        1827 & 5:38:46.824 & -69:06:03.24 & 4.59 & 0.01 & 4.16 & 0.06 & 5.75 & 0.05 & 1.8 & 14.22 & 67 & 55 & 2.55 & 130 & O6 \\
                        1439 & 5:38:40.848 & -69:06:12.24 & 4.58 & 0.01 & 3.91 & 0.22 & 5.04 & 0.05 & 2.0 & 16.14 & 40 & 26 & 2.77 & 110 & O7 \\
                        2763 & 5:38:45 & -69:05:38.76 & 4.58 & 0.01 & 3.91 & 0.13 & 5.19 & 0.05 & 1.4 & 15.15 & 40 & 29 & 2.77 & 200 & O6 \\
                        2911 & 5:38:45.792 & -69:05:40.92 & 4.58 & 0.01 & 4.11 & 0.07 & 5.33 & 0.05 & 1.4 & 14.80 & 94 & 33 & 2.58 & 90 & O5 \\
                        1346 & 5:38:41.28 & -69:06:16.92 & 4.58 & 0.01 & 4.11 & 0.17 & 5.39 & 0.05 & 2.2 & 15.45 & 94 & 37 & 2.58 & 270 & O6 \\
                        1969 & 5:38:44.616 & -69:05:52.44 & 4.58 & 0.01 & 4.01 & 0.20 & 4.84 & 0.05 & 1.8 & 16.42 & 146 & 24 & 1.95 & 10 & O6 \\
                        1274 & 5:38:39.72 & -69:06:24.12 & 4.58 & 0.01 & 4.21 & 0.10 & 5.74 & 0.05 & 1.8 & 14.19 & 166 & 48 & 2.06 & 140 & O5 \\
                        1472 & 5:38:45.264 & -69:06:13.68 & 4.58 & 0.01 & 3.91 & 0.13 & 4.94 & 0.05 & 1.6 & 15.97 & 40 & 24 & 2.77 & 40 & O6 \\
                        1651 & 5:38:38.016 & -69:06:09.72 & 4.58 & 0.03 & 4.21 & 0.17 & 5.12 & 0.08 & 2.2 & 16.14 & 166 & 28 & 2.06 & 110 & O6 \\
                        1373 & 5:38:36.048 & -69:06:16.92 & 4.58 & 0.01 & 4.21 & 0.08 & 5.59 & 0.05 & 2.2 & 14.96 & 166 & 41 & 2.06 & 50 & O5 \\
                        1409 & 5:38:41.04 & -69:06:15.12 & 4.58 & 0.01 & 4.01 & 0.19 & 5.25 & 0.05 & 2.2 & 15.80 & 146 & 31 & 1.95 & 290 & O6 \\
                        2053 & 5:38:46.128 & -69:05:54.6 & 4.58 & 0.01 & 4.11 & 0.10 & 5.63 & 0.05 & 1.8 & 14.44 & 94 & 43 & 2.58 & 90 & O6 \\
                        689 & 5:38:47.736 & -69:06:45 & 4.57 & 0.01 & 3.87 & 0.13 & 4.90 & 0.05 & 1.6 & 16.01 & 33 & 23 & 3.56 & 160 & O7 \\
                        1415 & 5:38:43.392 & -69:06:11.88 & 4.57 & 0.01 & 4.07 & 0.13 & 4.63 & 0.05 & 1.4 & 16.49 & 53 & -- & 2.90 & 170 & O7 \\
                        2570 & 5:38:39.648 & -69:05:26.16 & 4.57 & 0.01 & 4.17 & 0.08 & 5.16 & 0.04 & 1.2 & 14.95 & 71 & 27 & 2.73 & 390 & O5 \\
                        2447 & 5:38:40.8 & -69:05:25.08 & 4.57 & 0.01 & 4.27 & 0.08 & 5.53 & 0.05 & 1.8 & 14.65 & 92 & 39 & 2.52 & 420 & O5 \\
                        1725 & 5:38:40.824 & -69:06:04.68 & 4.57 & 0.02 & 2.87 & 0.48 & 4.73 & 0.07 & 1.4 & 16.25 & -- & 22 & -- & 110 & O7 \\
                        1522 & 5:38:42.072 & -69:06:09 & 4.57 & 0.02 & 4.17 & 0.15 & 4.97 & 0.06 & 1.6 & 15.85 & 71 & 24 & 2.73 & 330 & O7 \\
                        1857 & 5:38:41.448 & -69:05:57.84 & 4.57 & 0.01 & 3.97 & 0.19 & 5.32 & 0.05 & 1.6 & 14.98 & 41 & 31 & 3.25 & 250 & O6 \\
                        3007 & 5:38:32.88 & -69:05:44.52 & 4.57 & 0.01 & 3.97 & 0.19 & 4.99 & 0.05 & 2.0 & 16.20 & 41 & 24 & 3.25 & 40 & O7 \\
                        195 & 5:38:41.04 & -69:07:00.12 & 4.57 & 0.01 & 3.87 & 0.20 & 4.99 & 0.05 & 2.4 & 16.58 & 33 & 24 & 3.56 & 240 & O6 \\
                        1325 & 5:38:46.104 & -69:06:15.48 & 4.57 & 0.01 & 3.87 & 0.16 & 5.22 & 0.05 & 1.6 & 15.20 & 33 & 28 & 3.56 & 220 & O7 \\
                        2016 & 5:38:41.232 & -69:05:52.08 & 4.56 & 0.01 & 4.22 & 0.13 & 5.16 & 0.05 & 1.8 & 15.51 & 127 & 27 & 2.13 & 410 & O6 \\
                        1943 & 5:38:44.352 & -69:05:54.6 & 4.56 & 0.01 & 4.12 & 0.08 & 5.93 & 0.05 & 2.0 & 13.79 & 94 & 70 & 2.92 & 170 & O5 \\
                        3030 & 5:38:42.144 & -69:05:45.24 & 4.56 & 0.06 & 4.32 & 0.05 & 5.15 & 0.15 & 1.6 & 15.33 & 151 & 27 & 1.73 & 360 & O4 \\
                        2189 & 5:38:40.584 & -69:05:42.72 & 4.56 & 0.02 & 4.22 & 0.26 & 4.97 & 0.06 & 1.6 & 15.77* & 127 & 24 & 2.13 & 160 & O7 \\
                        2977 & 5:38:44.088 & -69:05:44.52 & 4.56 & 0.01 & 4.12 & 0.11 & 5.31 & 0.05 & 1.4 & 14.72 & 94 & 30 & 2.92 & 220 & O7 \\
                        1445 & 5:38:39.72 & -69:06:12.24 & 4.56 & 0.01 & 4.02 & 0.20 & 4.83 & 0.05 & 1.6 & 16.13 & 88 & 22 & 2.84 & 290 & O6 \\
                        3081 & 5:38:40.392 & -69:05:43.8 & 4.56 & 0.01 & 4.22 & 0.07 & 5.28 & 0.05 & 1.8 & 15.21 & 127 & 29 & 2.13 & 220 & O7 \\
                        2782 & 5:38:46.968 & -69:05:36.96 & 4.56 & 0.01 & 3.72 & 0.27 & 4.82 & 0.05 & 1.4 & 15.95 & 27 & 22 & 3.94 & 110 & O8 \\
                        3200 & 5:38:44.592 & -69:05:12.12 & 4.56 & 0.01 & 3.52 & 0.27 & 4.82 & 0.05 & 1.8 & 16.36 & 23 & 22 & 2.50 & 10 & O7 \\
                        2913 & 5:38:44.184 & -69:05:42 & 4.56 & 0.01 & 4.02 & 0.09 & 5.37 & 0.05 & 1.4 & 14.59 & 88 & 31 & 2.84 & 380 & O7 \\
                        1896 & 5:38:44.52 & -69:05:55.68 & 4.56 & 0.02 & 4.22 & 0.10 & 5.42 & 0.06 & 1.8 & 14.84 & 127 & 33 & 2.13 & 280 & O5 \\
                        2385 & 5:38:38.832 & -69:05:25.44 & 4.56 & 0.02 & 4.22 & 0.11 & 5.12 & 0.06 & 2.0 & 15.81 & 127 & 26 & 2.13 & 590 & O6 \\
                        960 & 5:38:40.08 & -69:06:31.68 & 4.56 & 0.02 & 4.22 & 0.18 & 4.74 & 0.06 & 2.8 & 17.56 & 127 & 21 & 2.13 & 40 & O7 \\
                        1858 & 5:38:39.672 & -69:05:59.28 & 4.54 & 0.02 & 4.07 & 0.56 & 4.79 & 0.06 & 1.8 & 16.32 & 71 & 20 & 3.38 & 170 & O7 \\
                        1572 & 5:38:38.784 & -69:06:13.32 & 4.54 & 0.16 & 4.37 & 0.05 & 5.20 & 0.38 & 1.6 & 15.07 & 103 & 28 & 2.12 & 280 & O6 \\
                        1580 & 5:38:42.672 & -69:06:11.16 & 4.54 & 0.02 & 3.37 & 0.43 & 5.18 & 0.06 & 2.0 & 15.52 & 19 & 28 & 2.35 & 200 & O6 \\
                        466 & 5:38:37.056 & -69:06:50.76 & 4.54 & 0.01 & 4.27 & 0.12 & 5.18 & 0.05 & 2.0 & 15.53 & 98 & 28 & 2.22 & 370 & O7 \\
                        1714 & 5:38:39.84 & -69:06:05.4 & 4.54 & 0.02 & 4.27 & 0.19 & 4.73 & 0.06 & 1.4 & 16.06 & 98 & 20 & 2.22 & 370 & O7 \\
                        1864 & 5:38:34.656 & -69:06:05.76 & 4.54 & 0.01 & 3.87 & 0.30 & 5.19 & 0.05 & 2.4 & 15.90 & 32 & 28 & 3.94 & 120 & O7 \\
                        1870 & 5:38:46.92 & -69:05:58.56 & 4.54 & 0.01 & 3.97 & 0.23 & 4.72 & 0.05 & 1.6 & 16.27 & 43 & 19 & 3.45 & 190 & O7 \\
                        1371 & 5:38:41.736 & -69:06:19.08 & 4.54 & 0.01 & 4.27 & 0.07 & 5.63 & 0.05 & 2.0 & 14.40 & 98 & 43 & 2.22 & 360 & O6 \\
                        290 & 5:38:39 & -69:06:59.04 & 4.54 & 0.01 & 4.17 & 0.10 & 5.37 & 0.05 & 2.6 & 15.66 & 96 & 31 & 2.80 & 200 & O7 \\
                        1998 & 5:38:38.712 & -69:05:54.96 & 4.54 & 0.01 & 4.17 & 0.36 & 4.65 & 0.05 & 2.4 & 17.25 & 96 & 19 & 2.80 & 50 & O8 \\
                        1527 & 5:38:42.456 & -69:06:09.36 & 4.54 & 0.02 & 4.07 & 0.31 & 5.16 & 0.06 & 2.0 & 15.57 & 71 & 27 & 3.38 & 210 & O7 \\
                        1941 & 5:38:36.84 & -69:05:56.4 & 4.54 & 0.01 & 3.97 & 0.18 & 4.94 & 0.05 & 1.6 & 15.73 & 43 & 23 & 3.45 & 170 & O6 \\
                        2005 & 5:38:47.184 & -69:05:54.6 & 4.54 & 0.01 & 3.77 & 0.15 & 4.91 & 0.04 & 1.6 & 15.80 & 28 & 22 & 4.02 & 160 & O7 \\
                        2944 & 5:38:44.544 & -69:05:41.28 & 4.54 & 0.02 & 3.87 & 0.39 & 4.69 & 0.06 & 1.6 & 16.35 & 32 & 19 & 3.94 & 180 & O7 \\
                        2193 & 5:38:45.264 & -69:05:46.32 & 4.54 & 0.04 & 4.27 & 0.05 & 5.61 & 0.10 & 1.4 & 13.85 & 98 & 41 & 2.22 & 310 & O6 \\
                        2223 & 5:38:37.344 & -69:05:21.48 & 4.54 & 0.01 & 3.77 & 0.28 & 4.46 & 0.05 & 1.0 & 16.32 & 28 & -- & 4.02 & 90 & O7 \\
                        2594 & 5:38:39.624 & -69:05:28.32 & 4.54 & 0.03 & 4.37 & 0.30 & 4.54 & 0.08 & 1.6 & 16.72 & 103 & 18 & 2.12 & 50 & O7 \\
                        2417 & 5:38:37.728 & -69:05:21.12 & 4.54 & 0.03 & 4.27 & 0.05 & 5.62 & 0.08 & 1.4 & 13.83 & 98 & 42 & 2.22 & 110 & O6 \\
                        2102 & 5:38:42.144 & -69:05:55.32 & 4.54 & 0.01 & 4.37 & 0.08 & 6.01 & 0.05 & 1.4 & 12.86 & 103 & 66 & 2.12 & 210 & O4 \\
                        2128 & 5:38:43.368 & -69:05:47.76 & 4.54 & 0.02 & 4.37 & 0.08 & 4.82 & 0.06 & 1.2 & 15.63 & 103 & 20 & 2.12 & 150 & O5 \\
                        2789 & 5:38:39.792 & -69:05:39.12 & 4.54 & 0.01 & 4.27 & 0.08 & 5.15 & 0.05 & 2.2 & 15.81 & 98 & 27 & 2.22 & 170 & O6 \\
                        2057 & 5:38:51.024 & -69:05:54.96 & 4.54 & 0.01 & 4.27 & 0.05 & 5.24 & 0.05 & 1.4 & 14.77 & 98 & 29 & 2.22 & 430 & O6 \\
                        2454 & 5:38:43.752 & -69:05:21.84 & 4.54 & 0.01 & 3.57 & 0.32 & 4.76 & 0.05 & 1.6 & 16.19 & 22 & 20 & 4.10 & 310 & O7 \\
                        1979 & 5:38:51.216 & -69:05:59.28 & 4.53 & 0.01 & 4.32 & 0.06 & 5.34 & 0.05 & 1.2 & 14.28 & 130 & 31 & 1.88 & 340 & O6 \\
                        2165 & 5:38:46.2 & -69:05:51.36 & 4.53 & 0.01 & 4.22 & 0.05 & 5.64 & 0.05 & 1.4 & 13.71 & 105 & 41 & 2.90 & 260 & O7 \\
                        2897 & 5:38:35.928 & -69:05:35.16 & 4.53 & 0.01 & 3.92 & 0.16 & 5.04 & 0.05 & 1.8 & 15.62 & 50 & 23 & 4.13 & 170 & O7 \\
                        774 & 5:38:52.728 & -69:06:43.2 & 4.53 & 0.07 & 4.32 & 0.05 & 5.50 & 0.17 & 1.6 & 14.27 & 130 & 36 & 1.88 & 110 & O6 \\
                        375 & 5:38:46.08 & -69:06:56.16 & 4.53 & 0.01 & 3.92 & 0.13 & 4.83 & 0.05 & 1.2 & 15.56 & 50 & 20 & 4.13 & 210 & O8 \\
                        2326 & 5:38:38.04 & -69:05:20.76 & 4.53 & 0.01 & 4.32 & 0.24 & 4.82 & 0.05 & 0.8 & 15.17* & 130 & 20 & 1.88 & 110 & O7 \\
                        1988 & 5:38:38.328 & -69:05:57.12 & 4.53 & 0.02 & 3.62 & 0.35 & 4.47 & 0.06 & 2.0 & 17.25 & 22 & 17 & 5.02 & 100 & O8 \\
                        2453 & 5:38:40.152 & -69:05:20.76 & 4.53 & 0.02 & 4.02 & 0.26 & 4.54 & 0.06 & 1.4 & 16.46 & 69 & 18 & 3.55 & 190 & O8 \\
                        1974 & 5:38:40.536 & -69:05:53.52 & 4.53 & 0.01 & 4.32 & 0.05 & 5.13 & 0.05 & 1.6 & 15.19 & 130 & 25 & 1.88 & 70 & O6 \\
                        2945 & 5:38:46.512 & -69:05:42 & 4.53 & 0.01 & 3.92 & 0.15 & 4.98 & 0.05 & 1.6 & 15.57 & 50 & 22 & 4.13 & 180 & O8 \\
                        2884 & 5:38:46.584 & -69:05:37.32 & 4.53 & 0.01 & 4.12 & 0.16 & 5.04 & 0.05 & 1.2 & 15.02 & 88 & 23 & 3.12 & 110 & O7 \\
                        3172 & 5:38:46.896 & -69:05:20.4 & 4.52 & 0.01 & 3.57 & 0.34 & 4.40 & 0.05 & 1.4 & 16.76 & 20 & 16 & 5.61 & 10 & O8 \\
                        2985 & 5:38:44.712 & -69:05:40.92 & 4.52 & 0.01 & 3.37 & 0.26 & 4.38 & 0.05 & 1.8 & 17.20 & 18 & 16 & 3.41 & 120 & O9 \\
                        3010 & 5:38:51.168 & -69:05:42 & 4.52 & 0.01 & 3.77 & 0.15 & 4.79 & 0.05 & 1.4 & 15.78 & 26 & 19 & 4.94 & 320 & O8 \\
                        2846 & 5:38:43.008 & -69:05:33.72 & 4.52 & 0.02 & 3.27 & 0.35 & 4.67 & 0.06 & 2.0 & 16.68 & 17 & 18 & 2.23 & 190 & O9 \\
                        3043 & 5:38:44.232 & -69:05:47.04 & 4.52 & 0.04 & 4.37 & 0.05 & 5.78 & 0.10 & 1.4 & 13.30 & 212 & 49 & 2.19 & 130 & O7 \\
                        3167 & 5:38:41.808 & -69:05:07.44 & 4.52 & 0.01 & 3.67 & 0.19 & 4.68 & 0.05 & 1.4 & 16.05 & 23 & 18 & 5.35 & 200 & O8 \\
                        2665 & 5:38:42.864 & -69:05:30.48 & 4.52 & 0.01 & 4.17 & 0.09 & 4.94 & 0.05 & 1.6 & 15.60 & 80 & 21 & 2.86 & 140 & O7 \\
                        543 & 5:38:40.08 & -69:06:46.8 & 4.52 & 0.02 & 3.97 & 0.36 & 4.67 & 0.06 & 2.8 & 17.48 & 36 & 18 & 4.12 & 200 & O7 \\
                        721 & 5:38:41.04 & -69:06:36.36 & 4.52 & 0.02 & 3.17 & 0.31 & 2.75 & 0.06 & 2.0 & 21.49* & 16 & -- & 0.90 & 70 & O9 \\
                        250 & 5:38:38.88 & -69:06:56.88 & 4.52 & 0.04 & 2.67 & 0.35 & 4.20 & 0.10 & 2.4 & 18.25 & -- & -- & -- & 10 & O8 \\
                        2908 & 5:38:41.72 & -69:05:05.9 & 4.52 & 0.02 & 3.37 & 0.34 & 4.94 & 0.06 & 1.4 & 15.42* & 18 & 21 & 3.41 & 260 & B0 \\
                        2521 & 5:38:43.224 & -69:05:27.6 & 4.52 & 0.01 & 3.57 & 0.16 & 4.87 & 0.04 & 1.4 & 15.59 & 20 & 20 & 5.61 & 140 & O8 \\
                        2714 & 5:38:41.544 & -69:05:34.8 & 4.52 & 0.02 & 3.47 & 0.40 & 4.46 & 0.06 & 1.4 & 16.61 & 19 & 17 & 4.56 & 120 & O9 \\
                        911 & 5:38:41.184 & -69:06:35.28 & 4.52 & 0.01 & 3.57 & 0.19 & 4.81 & 0.05 & 2.0 & 16.33 & 20 & 19 & 5.61 & 130 & O8 \\
                        2233 & 5:38:40.512 & -69:05:12.48 & 4.52 & 0.01 & 3.57 & 0.18 & 4.95 & 0.04 & 1.2 & 15.19 & 20 & 21 & 5.61 & 240 & O8 \\
                        2174 & 5:38:35.016 & -69:05:49.92 & 4.52 & 0.02 & 3.07 & 0.31 & 4.41 & 0.06 & 2.0 & 17.34 & -- & 16 & -- & 10 & O9 \\
                        1846 & 5:38:38.64 & -69:06:01.08 & 4.52 & 0.02 & 3.47 & 0.35 & 5.06 & 0.06 & 2.4 & 16.12* & 19 & 23 & 4.56 & 180 & O8 \\
                        1778 & 5:38:41.232 & -69:06:06.84 & 4.52 & 0.01 & 3.67 & 0.18 & 5.15 & 0.05 & 1.8 & 15.28 & 23 & 25 & 5.35 & 250 & O8 \\
                        1795 & 5:38:50.856 & -69:05:59.28 & 4.52 & 0.02 & 4.37 & 0.22 & 4.01 & 0.06 & 0.4 & 16.73* & 212 & -- & 2.19 & 170 & O7 \\
                        2302 & 5:38:34.056 & -69:05:15.36 & 4.52 & 0.01 & 3.77 & 0.22 & 4.51 & 0.04 & 1.6 & 16.67 & 26 & 17 & 4.94 & 210 & O8 \\
                        1275 & 5:38:43.824 & -69:06:21.96 & 4.52 & 0.01 & 3.47 & 0.22 & 4.68 & 0.04 & 1.6 & 16.27 & 19 & 18 & 4.56 & 180 & O8 \\
                        2301 & 5:38:44.376 & -69:05:14.64 & 4.52 & 0.01 & 3.57 & 0.23 & 4.90 & 0.05 & 1.6 & 15.70 & 20 & 20 & 5.61 & 180 & O8 \\
                        1641 & 5:38:46.848 & -69:06:06.48 & 4.52 & 0.02 & 2.97 & 0.33 & 4.70 & 0.06 & 2.6 & 17.22 & -- & 19 & -- & 110 & O9 \\
                        2033 & 5:38:45.912 & -69:05:50.64 & 4.52 & 0.01 & 4.17 & 0.11 & 4.91 & 0.05 & 1.4 & 15.48 & 80 & 21 & 2.86 & 390 & O7 \\
                        1207 & 5:38:43.08 & -69:06:23.04 & 4.52 & 0.01 & 3.77 & 0.28 & 4.70 & 0.05 & 1.8 & 16.40 & 26 & 19 & 4.94 & 220 & O8 \\
                        2498 & 5:38:40.224 & -69:05:28.68 & 4.52 & 0.01 & 3.57 & 0.24 & 4.90 & 0.05 & 1.8 & 15.92 & 20 & 20 & 5.61 & 260 & O8 \\
                        1297 & 5:38:41.04 & -69:06:16.92 & 4.52 & 0.02 & 4.17 & 0.34 & 4.11 & 0.06 & 2.2 & 18.28 & 80 & -- & 2.86 & 240 & O8 \\
                        1625 & 5:38:46.68 & -69:06:07.2 & 4.51 & 0.02 & 3.61 & 0.27 & 4.70 & 0.06 & 2.6 & 17.14 & 22 & 18 & 5.87 & 160 & O8 \\
                        1340 & 5:38:43.728 & -69:06:16.92 & 4.51 & 0.02 & 3.71 & 0.34 & 4.23 & 0.06 & 1.2 & 16.91 & 27 & -- & 5.34 & 210 & O8 \\
                        2256 & 5:38:33 & -69:05:13.2 & 4.51 & 0.01 & 3.91 & 0.18 & 4.88 & 0.05 & 1.4 & 15.49 & 36 & 20 & 4.29 & 240 & O8 \\
                        2389 & 5:38:41.664 & -69:05:23.64 & 4.51 & 0.01 & 3.91 & 0.22 & 4.78 & 0.04 & 1.6 & 15.95 & 36 & 19 & 4.29 & 250 & O9 \\
                        1387 & 5:38:45.696 & -69:06:15.48 & 4.51 & 0.02 & 3.41 & 0.26 & 4.55 & 0.06 & 1.6 & 16.52 & 18 & 17 & 4.24 & 200 & O8 \\
                        1334 & 5:38:49.056 & -69:06:19.8 & 4.51 & 0.01 & 3.91 & 0.15 & 4.88 & 0.05 & 1.4 & 15.51 & 36 & 20 & 4.29 & 120 & O8 \\
                        2388 & 5:38:43.056 & -69:05:26.52 & 4.51 & 0.01 & 3.51 & 0.21 & 4.53 & 0.04 & 1.6 & 16.58 & 19 & 17 & 5.38 & 220 & O9 \\
                        2545 & 5:38:42.648 & -69:05:22.2 & 4.51 & 0.01 & 3.61 & 0.18 & 4.96 & 0.05 & 1.4 & 15.30 & 22 & 21 & 5.87 & 260 & O8 \\
                        2098 & 5:38:33.888 & -69:05:51.72 & 4.51 & 0.02 & 3.81 & 0.32 & 4.50 & 0.06 & 1.8 & 16.86 & 31 & 17 & 4.81 & 280 & O8 \\
                        1040 & 5:38:44.568 & -69:06:28.8 & 4.51 & 0.01 & 3.51 & 0.17 & 4.86 & 0.05 & 1.8 & 15.94 & 19 & 20 & 5.38 & 230 & O9 \\
                        3180 & 5:38:45.168 & -69:05:08.52 & 4.51 & 0.01 & 4.41 & 0.05 & 5.06 & 0.05 & 1.6 & 15.25 & 163 & 23 & 2.18 & 220 & O7 \\
                        1177 & 5:38:39.144 & -69:06:25.2 & 4.51 & 0.01 & 4.01 & 0.19 & 5.17 & 0.05 & 2.2 & 15.58 & 41 & 27 & 3.76 & 200 & O8 \\
                        3112 & 5:38:42 & -69:05:13.2 & 4.51 & 0.01 & 4.41 & 0.05 & 5.13 & 0.05 & 1.2 & 14.66 & 163 & 26 & 2.18 & 10 & O6 \\
                        2038 & 5:38:50.592 & -69:05:54.6 & 4.51 & 0.01 & 3.81 & 0.16 & 4.61 & 0.05 & 1.4 & 16.18 & 31 & 18 & 4.81 & 200 & O8 \\
                        3027 & 5:38:51.792 & -69:05:47.04 & 4.51 & 0.01 & 3.61 & 0.15 & 4.71 & 0.05 & 1.4 & 15.93 & 22 & 18 & 5.87 & 350 & O8 \\
                        2876 & 5:38:46.536 & -69:05:44.16 & 4.51 & 0.01 & 3.71 & 0.15 & 5.08 & 0.04 & 1.8 & 15.39 & 27 & 24 & 5.34 & 230 & O8 \\
                        2231 & 5:38:41.088 & -69:05:13.2 & 4.51 & 0.05 & 4.41 & 0.05 & 4.93 & 0.12 & 1.0 & 14.96 & 163 & 21 & 2.18 & 170 & O6 \\
                        3149 & 5:38:38.928 & -69:05:12.12 & 4.51 & 0.02 & 3.31 & 0.34 & 4.31 & 0.06 & 1.2 & 16.73 & 17 & 15 & 3.10 & 150 & B1 \\
                        710 & 5:38:44.376 & -69:06:41.04 & 4.49 & 0.01 & 3.76 & 0.25 & 4.47 & 0.05 & 1.6 & 16.59 & 29 & 15 & 5.08 & 150 & O8 \\
                        1561 & 5:38:42.336 & -69:06:05.04 & 4.49 & 0.03 & 4.46 & 0.05 & 6.18 & 0.08 & 1.4 & 12.13* & 154 & 85 & 1.75 & 10 & O7 \\
                        1223 & 5:38:45.912 & -69:06:20.88 & 4.49 & 0.01 & 4.36 & 0.06 & 5.43 & 0.05 & 1.6 & 14.20 & 116 & 32 & 1.82 & 130 & O7 \\
                        2270 & 5:38:42.864 & -69:05:15 & 4.49 & 0.01 & 3.56 & 0.19 & 4.84 & 0.05 & 1.4 & 15.49 & 20 & 20 & 5.81 & 190 & O9 \\
                        3106 & 5:38:39.048 & -69:05:45.96 & 4.49 & 0.03 & 3.86 & 0.43 & 4.50 & 0.08 & 2.8 & 17.73 & 28 & 15 & 5.33 & 10 & O8 \\
                        2815 & 5:38:47.904 & -69:05:33 & 4.49 & 0.02 & 3.56 & 0.28 & 4.58 & 0.06 & 2.0 & 16.73 & 20 & 16 & 5.81 & 220 & O9 \\
                        909 & 5:38:43.128 & -69:06:37.08 & 4.49 & 0.02 & 3.56 & 0.28 & 4.60 & 0.06 & 2.0 & 16.69 & 20 & 16 & 5.81 & 140 & O9 \\
                        813 & 5:38:50.688 & -69:06:37.8 & 4.49 & 0.01 & 3.96 & 0.21 & 4.48 & 0.04 & 1.8 & 16.77 & 32 & 15 & 4.65 & 160 & O8 \\
                        2735 & 5:38:44.4 & -69:05:33.36 & 4.49 & 0.02 & 3.86 & 0.31 & 4.11 & 0.06 & 1.2 & 17.10 & 28 & -- & 5.33 & 290 & O8 \\
                        1398 & 5:38:47.376 & -69:06:18 & 4.49 & 0.01 & 3.86 & 0.19 & 4.66 & 0.05 & 1.4 & 15.93 & 28 & 17 & 5.33 & 430 & O8 \\
                        2130 & 5:38:42.096 & -69:05:52.08 & 4.49 & 0.02 & 4.36 & 0.53 & 4.68 & 0.06 & 1.8 & 16.28 & 116 & 17 & 1.82 & 10 & O8 \\
                        2848 & 5:38:50.424 & -69:05:34.8 & 4.48 & 0.04 & 3.30 & 0.33 & 3.88 & 0.10 & 1.4 & 17.81 & 15 & -- & 7.21 & 140 & B0 \\
                        2771 & 5:38:46.752 & -69:05:39.12 & 4.48 & 0.03 & 3.50 & 0.25 & 4.14 & 0.08 & 1.2 & 16.98 & 20 & 13 & 6.65 & 160 & O9 \\
                        2097 & 5:38:37.992 & -69:05:49.92 & 4.48 & 0.03 & 3.70 & 0.25 & 4.41 & 0.08 & 1.6 & 16.70 & 33 & 14 & 5.48 & 370 & O9 \\
                        2090 & 5:38:31.416 & -69:05:53.16 & 4.48 & 0.01 & 3.90 & 0.12 & 4.51 & 0.05 & 1.6 & 16.45 & 29 & 15 & 5.08 & 270 & O9 \\
                        2401 & 5:38:45.408 & -69:05:24.72 & 4.48 & 0.03 & 3.30 & 0.24 & 4.31 & 0.08 & 1.2 & 16.53 & 15 & 14 & 7.21 & 170 & O9 \\
                        249 & 5:38:37.44 & -69:06:57.96 & 4.48 & 0.07 & 3.10 & 0.21 & 4.39 & 0.17 & 2.2 & 17.35 & 14 & 14 & 3.61 & 120 & B0 \\
                        225 & 5:38:37.776 & -69:06:59.4 & 4.48 & 0.01 & 4.40 & 0.14 & 4.57 & 0.05 & 2.4 & 17.10 & 267 & 16 & 2.14 & 100 & O8 \\
                        3151 & 5:38:40.584 & -69:05:08.88 & 4.48 & 0.05 & 3.20 & 0.32 & 4.08 & 0.12 & 1.2 & 17.13 & 14 & 13 & 5.42 & 230 & O9 \\
                        489 & 5:38:43.44 & -69:06:47.88 & 4.48 & 0.07 & 3.00 & 0.30 & 4.02 & 0.17 & 1.6 & 17.67 & 13 & -- & 1.67 & 180 & B0 \\
                        2114 & 5:38:46.752 & -69:05:48.84 & 4.48 & 0.02 & 3.70 & 0.21 & 4.44 & 0.06 & 1.4 & 16.43 & 33 & 15 & 5.48 & 130 & O9 \\
                        2511 & 5:38:32.808 & -69:05:25.08 & 4.48 & 0.02 & 3.60 & 0.24 & 4.40 & 0.06 & 1.4 & 16.51 & 23 & 14 & 6.19 & 200 & O9 \\
                        2086 & 5:38:37.872 & -69:05:48.48 & 4.48 & 0.10 & 3.00 & 0.42 & 4.06 & 0.23 & 1.4 & 17.38 & 13 & 13 & 1.67 & 100 & O9 \\
                        2607 & 5:38:33.552 & -69:05:21.84 & 4.48 & 0.02 & 3.60 & 0.19 & 4.53 & 0.06 & 1.4 & 16.19 & 23 & 15 & 6.19 & 150 & B0 \\
                        2776 & 5:38:43.152 & -69:05:37.68 & 4.48 & 0.02 & 3.90 & 0.29 & 4.75 & 0.06 & 1.6 & 15.85 & 29 & 18 & 5.08 & 120 & O8 \\
                        2266 & 5:38:44.736 & -69:05:14.28 & 4.48 & 0.03 & 3.90 & 0.41 & 4.40 & 0.08 & 1.6 & 16.73 & 29 & 14 & 5.08 & 140 & O7 \\
                        2718 & 5:38:46.416 & -69:05:30.84 & 4.48 & 0.01 & 3.80 & 0.12 & 4.94 & 0.05 & 1.4 & 15.18 & 34 & 21 & 5.07 & 180 & O9 \\
                        835 & 5:38:48.312 & -69:06:34.56 & 4.48 & 0.04 & 3.00 & 0.32 & 4.13 & 0.10 & 1.8 & 17.60 & 13 & 13 & 1.67 & 120 & B0 \\
                        2176 & 5:38:31.704 & -69:05:49.2 & 4.48 & 0.02 & 3.20 & 0.20 & 4.42 & 0.06 & 1.4 & 16.46 & 14 & 15 & 5.42 & 110 & B0 \\
                        2150 & 5:38:37.152 & -69:05:48.84 & 4.48 & 0.04 & 3.40 & 0.35 & 4.20 & 0.10 & 1.6 & 17.22 & 16 & 13 & 7.33 & 180 & B0 \\
                        2653 & 5:38:41.328 & -69:05:32.28 & 4.48 & 0.01 & 4.20 & 0.05 & 5.62 & 0.04 & 1.6 & 13.68 & 168 & 38 & 3.37 & 200 & O8 \\
                        2703 & 5:38:40.272 & -69:05:31.2 & 4.48 & 0.03 & 3.70 & 0.31 & 4.60 & 0.08 & 1.6 & 16.21 & 33 & 16 & 5.48 & 200 & O8 \\
                        1370 & 5:38:51.384 & -69:06:17.28 & 4.48 & 0.02 & 3.50 & 0.17 & 4.31 & 0.06 & 1.4 & 16.74 & 20 & 14 & 6.65 & 240 & O9 \\
                        1428 & 5:38:40.128 & -69:06:14.04 & 4.48 & 0.04 & 3.50 & 0.32 & 4.42 & 0.10 & 2.2 & 17.26 & 20 & 15 & 6.65 & 170 & B0 \\
                        123 & 5:38:43.128 & -69:07:01.92 & 4.48 & 0.02 & 4.00 & 0.29 & 4.16 & 0.06 & 1.2 & 16.92 & 35 & 13 & 4.32 & 120 & O8 \\
                        1138 & 5:38:41.016 & -69:06:25.56 & 4.48 & 0.02 & 3.70 & 0.23 & 4.57 & 0.06 & 2.2 & 16.89 & 33 & 16 & 5.48 & 260 & O9 \\
                        70 & 5:38:41.547 & -69:07:02.111 & 4.48 & 0.02 & 4.20 & 0.45 & 4.44 & 0.06 & 1.2 & 16.21* & 168 & 15 & 3.37 & 170 & O8 \\
                        1968 & 5:38:31.416 & -69:05:57.12 & 4.48 & 0.07 & 3.30 & 0.24 & 4.38 & 0.17 & 1.6 & 16.77 & 15 & 14 & 7.21 & 220 & O9 \\
                        2939 & 5:38:43.824 & -69:05:38.76 & 4.48 & 0.03 & 3.60 & 0.33 & 4.26 & 0.08 & 1.2 & 16.68 & 23 & 14 & 6.19 & 120 & O9 \\
                        2966 & 5:38:37.104 & -69:05:40.2 & 4.48 & 0.02 & 3.70 & 0.21 & 4.50 & 0.06 & 1.6 & 16.46 & 33 & 15 & 5.48 & 200 & O9 \\
                        1226 & 5:38:44.88 & -69:06:18.72 & 4.48 & 0.04 & 3.20 & 0.25 & 3.87 & 0.10 & 1.4 & 17.84 & 14 & -- & 5.42 & 210 & O9 \\
                        1143 & 5:38:43.584 & -69:06:24.12 & 4.48 & 0.04 & 3.40 & 0.30 & 4.45 & 0.10 & 1.8 & 16.80 & 16 & 15 & 7.33 & 290 & B0 \\
                        935 & 5:38:41.256 & -69:06:32.4 & 4.48 & 0.02 & 3.90 & 0.23 & 4.74 & 0.06 & 2.6 & 16.88 & 29 & 18 & 5.08 & 220 & O9 \\
                        1457 & 5:38:34.488 & -69:06:14.76 & 4.48 & 0.04 & 3.00 & 0.25 & 4.53 & 0.10 & 1.8 & 16.60 & 13 & 15 & 1.67 & 130 & B0 \\
                        3032 & 5:38:49.08 & -69:05:47.4 & 4.48 & 0.07 & 3.20 & 0.22 & 4.05 & 0.16 & 1.6 & 17.59 & 14 & 13 & 5.42 & 150 & O9 \\
                        1973 & 5:38:39.816 & -69:05:54.24 & 4.48 & 0.04 & 3.30 & 0.38 & 4.63 & 0.10 & 2.2 & 16.75 & 15 & 16 & 7.21 & 100 & B1 \\
                        3062 & 5:38:43.104 & -69:05:40.2 & 4.48 & 0.01 & 4.20 & 0.05 & 5.60 & 0.05 & 1.6 & 13.72 & 168 & 37 & 3.37 & 190 & O8 \\
                        898 & 5:38:49.416 & -69:06:32.4 & 4.48 & 0.09 & 2.60 & 0.38 & 3.98 & 0.21 & 1.6 & 17.77 & -- & -- & -- & 80 & B0 \\
                        1456 & 5:38:45.96 & -69:06:12.6 & 4.48 & 0.04 & 3.50 & 0.30 & 4.27 & 0.10 & 1.6 & 17.05 & 20 & 14 & 6.65 & 170 & O9 \\
                        1464 & 5:38:48.144 & -69:06:12.6 & 4.48 & 0.02 & 3.60 & 0.21 & 4.29 & 0.06 & 1.2 & 16.59 & 23 & 14 & 6.19 & 130 & O9 \\
                        520 & 5:38:45.408 & -69:06:48.24 & 4.48 & 0.04 & 3.30 & 0.27 & 4.16 & 0.10 & 1.8 & 17.52 & 15 & 13 & 7.21 & 200 & O9 \\
                        1826 & 5:38:48.168 & -69:06:05.04 & 4.48 & 0.02 & 3.50 & 0.21 & 4.35 & 0.06 & 1.6 & 16.85 & 20 & 14 & 6.65 & 230 & O9 \\
                        1847 & 5:38:44.952 & -69:06:01.08 & 4.48 & 0.03 & 3.40 & 0.34 & 4.19 & 0.08 & 1.8 & 17.45 & 16 & 13 & 7.33 & 80 & B0 \\
                        1505 & 5:38:41.4 & -69:06:09.36 & 4.48 & 0.03 & 4.20 & 0.50 & 4.13 & 0.08 & 1.4 & 17.20 & 168 & 13 & 3.37 & 210 & O8 \\
                        912 & 5:38:47.208 & -69:06:31.32 & 4.48 & 0.03 & 3.50 & 0.35 & 4.35 & 0.08 & 1.6 & 16.84 & 20 & 14 & 6.65 & 100 & O9 \\
                        410 & 5:38:49.68 & -69:06:55.08 & 4.46 & 0.02 & 3.54 & 0.21 & 4.28 & 0.06 & 1.8 & 17.11 & 17 & 13 & 8.22 & 130 & O9 \\
                        1632 & 5:38:36.312 & -69:06:08.28 & 4.46 & 0.01 & 4.44 & 0.05 & 5.17 & 0.05 & 1.4 & 14.48 & 180 & 23 & 2.21 & 10 & O8 \\
                        2681 & 5:38:43.584 & -69:05:29.4 & 4.46 & 0.05 & 3.44 & 0.29 & 4.50 & 0.13 & 1.4 & 16.16 & 17 & 16 & 5.77 & 290 & B0 \\
                        1293 & 5:38:38.52 & -69:06:21.96 & 4.46 & 0.02 & 4.44 & 0.05 & 5.69 & 0.06 & 1.6 & 13.39 & 180 & 41 & 2.21 & 140 & O7 \\
                        2813 & 5:38:33.192 & -69:05:29.04 & 4.46 & 0.05 & 3.44 & 0.28 & 4.10 & 0.13 & 1.4 & 17.15 & 17 & 12 & 5.77 & 110 & O9 \\
                        1083 & 5:38:42.216 & -69:06:24.12 & 4.46 & 0.02 & 4.04 & 0.15 & 5.01 & 0.06 & 1.4 & 14.88* & 66 & 20 & 4.52 & 190 & O9 \\
                        1901 & 5:38:45.432 & -69:06:00.36 & 4.45 & 0.05 & 3.48 & 0.23 & 4.43 & 0.12 & 1.6 & 16.47 & 18 & 14 & 7.91 & 160 & B1 \\
                        3018 & 5:38:38.376 & -69:05:08.16 & 4.45 & 0.08 & 3.38 & 0.32 & 3.96 & 0.18 & 1.2 & 17.23 & 17 & 11 & 7.26 & 130 & O9 \\
                        2024 & 5:38:44.232 & -69:05:51 & 4.45 & 0.05 & 3.28 & 0.22 & 4.44 & 0.12 & 1.4 & 16.24 & 14 & 14 & 8.96 & 250 & O9 \\
                        1282 & 5:38:42.552 & -69:06:19.44 & 4.45 & 0.04 & 3.68 & 0.24 & 4.04 & 0.10 & 1.6 & 17.44 & 25 & 12 & 7.29 & 290 & B0 \\
                        3129 & 5:38:52.296 & -69:05:44.88 & 4.45 & 0.08 & 3.48 & 0.26 & 3.94 & 0.18 & 1.8 & 17.88 & 18 & 11 & 7.91 & 170 & O9 \\
                        1152 & 5:38:46.848 & -69:06:26.64 & 4.45 & 0.03 & 3.58 & 0.19 & 4.70 & 0.08 & 1.8 & 16.00 & 19 & 16 & 8.13 & 190 & O9 \\
                        888 & 5:38:47.16 & -69:06:36 & 4.45 & 0.03 & 3.48 & 0.14 & 4.48 & 0.08 & 1.8 & 16.54 & 18 & 14 & 7.91 & 200 & B0 \\
                        2506 & 5:38:46.92 & -69:05:26.16 & 4.45 & 0.06 & 3.18 & 0.37 & 4.05 & 0.15 & 1.8 & 17.63 & 13 & 12 & 8.55 & 100 & B0 \\
                        2928 & 5:38:35.352 & -69:05:07.08 & 4.45 & 0.05 & 3.98 & 0.30 & 4.27 & 0.12 & 2.0 & 17.26 & 33 & 13 & 4.83 & 450 & O9 \\
                        920 & 5:38:46.368 & -69:06:32.4 & 4.45 & 0.09 & 3.28 & 0.48 & 3.88 & 0.21 & 2.2 & 18.44 & 14 & 11 & 8.96 & 450 & B0 \\
                        2018 & 5:38:44.4 & -69:05:49.92 & 4.45 & 0.08 & 3.08 & 0.28 & 4.45 & 0.18 & 1.6 & 16.41 & 12 & 14 & 8.13 & 230 & B0 \\
                        1062 & 5:38:42.432 & -69:06:24.12 & 4.45 & 0.03 & 4.18 & 0.29 & 4.67 & 0.08 & 1.2 & 15.47* & 65 & 16 & 3.61 & 70 & O9 \\
                        1894 & 5:38:51.768 & -69:06:01.44 & 4.45 & 0.02 & 3.68 & 0.14 & 4.47 & 0.06 & 1.2 & 15.96 & 25 & 14 & 7.29 & 250 & B0 \\
                        178 & 5:38:48.648 & -69:07:00.12 & 4.45 & 0.08 & 3.28 & 0.38 & 3.89 & 0.18 & 1.8 & 18.02 & 14 & 11 & 8.96 & 200 & O9 \\
                        1390 & 5:38:44.928 & -69:06:17.64 & 4.45 & 0.04 & 3.68 & 0.23 & 4.42 & 0.10 & 1.4 & 16.29 & 25 & 14 & 7.29 & 280 & O9 \\
                        2159 & 5:38:47.952 & -69:05:47.04 & 4.45 & 0.08 & 3.28 & 0.34 & 4.14 & 0.19 & 2.4 & 18.00 & 14 & 12 & 8.96 & 150 & B1 \\
                        1210 & 5:38:47.784 & -69:06:21.96 & 4.45 & 0.04 & 3.58 & 0.19 & 4.17 & 0.10 & 1.4 & 16.91 & 19 & 12 & 8.13 & 190 & O9 \\
                        501 & 5:38:43.608 & -69:06:51.48 & 4.45 & 0.03 & 3.68 & 0.21 & 4.31 & 0.08 & 1.6 & 16.77 & 25 & 13 & 7.29 & 240 & O9 \\
                        1276 & 5:38:42.648 & -69:06:18 & 4.45 & 0.10 & 3.58 & 0.35 & 4.39 & 0.23 & 1.6 & 16.58 & 19 & 14 & 8.13 & 310 & B0 \\
                        1434 & 5:38:49.44 & -69:06:15.48 & 4.45 & 0.02 & 3.38 & 0.15 & 4.19 & 0.06 & 1.2 & 16.67 & 17 & 12 & 7.26 & 290 & B0 \\
                        1453 & 5:38:44.448 & -69:06:11.88 & 4.45 & 0.08 & 3.38 & 0.38 & 4.22 & 0.19 & 1.6 & 17.00 & 17 & 12 & 7.26 & 230 & B0 \\
                        2543 & 5:38:31.656 & -69:05:27.24 & 4.45 & 0.05 & 3.48 & 0.18 & 4.40 & 0.12 & 1.4 & 16.35 & 18 & 14 & 7.91 & 260 & B0 \\
                        2140 & 5:38:47.808 & -69:05:50.28 & 4.45 & 0.05 & 3.38 & 0.26 & 4.03 & 0.12 & 1.6 & 17.46 & 17 & 12 & 7.26 & 160 & B0 \\
                        367 & 5:38:41.976 & -69:06:53.64 & 4.45 & 0.03 & 3.28 & 0.12 & 4.36 & 0.08 & 1.6 & 16.64 & 14 & 13 & 8.96 & 220 & B0 \\
                        805 & 5:38:48.456 & -69:06:36.72 & 4.45 & 0.06 & 3.38 & 0.22 & 4.27 & 0.14 & 1.4 & 16.67 & 17 & 13 & 7.26 & 190 & B0 \\
                        1504 & 5:38:39.096 & -69:06:10.44 & 4.45 & 0.08 & 3.68 & 0.39 & 4.61 & 0.18 & 2.4 & 16.81 & 25 & 15 & 7.29 & 220 & B1 \\
                        2147 & 5:38:46.656 & -69:05:46.68 & 4.45 & 0.07 & 3.68 & 0.41 & 4.22 & 0.16 & 2.0 & 17.38 & 25 & 12 & 7.29 & 120 & O9 \\
                        1184 & 5:38:42.24 & -69:06:25.92 & 4.41 & 0.01 & 4.35 & 0.08 & 5.42 & 0.05 & 1.6 & 13.77 & 85 & 29 & 3.24 & 170 & O9 \\
                        1764 & 5:38:46.128 & -69:06:03.96 & 4.41 & 0.05 & 3.65 & 0.27 & 3.97 & 0.12 & 1.8 & 17.58 & 18 & 10 & 9.12 & 240 & O9 \\
                        1607 & 5:38:38.952 & -69:06:06.84 & 4.41 & 0.10 & 3.35 & 0.45 & 4.12 & 0.23 & 1.6 & 17.01 & 15 & 11 & 10.27 & 220 & O9 \\
                        2804 & 5:38:44.952 & -69:05:33 & 4.41 & 0.02 & 3.75 & 0.12 & 4.70 & 0.06 & 1.2 & 15.15 & 21 & 16 & 8.02 & 130 & B0 \\
                        2610 & 5:38:43.968 & -69:05:28.32 & 4.41 & 0.09 & 3.55 & 0.42 & 4.25 & 0.22 & 1.4 & 16.48 & 16 & 12 & 10.22 & 270 & B0 \\
                        1221 & 5:38:40.824 & -69:06:21.24 & 4.41 & 0.08 & 3.65 & 0.35 & 4.23 & 0.19 & 2.8 & 17.92 & 18 & 12 & 9.12 & 370 & O9 \\
                        549 & 5:38:41.832 & -69:06:48.96 & 4.41 & 0.08 & 3.75 & 0.22 & 4.25 & 0.19 & 1.6 & 16.68 & 21 & 12 & 8.02 & 530 & B0 \\
                        1565 & 5:38:49.968 & -69:06:04.32 & 4.41 & 0.09 & 4.35 & 0.34 & 3.83 & 0.20 & 0.6 & 16.74* & 85 & 10 & 3.24 & 330 & B0 \\
                        1768 & 5:38:46.08 & -69:06:02.52 & 4.41 & 0.09 & 3.95 & 0.33 & 3.18 & 0.22 & 1.6 & 19.37 & 29 & -- & 6.02 & 270 & B0 \\
                        1405 & 5:38:51.048 & -69:06:20.52 & 4.41 & 0.01 & 4.45 & 0.05 & 5.26 & 0.05 & 1.4 & 13.95 & 102 & 26 & 2.53 & 150 & O9 \\
                        2929 & 5:38:49.632 & -69:05:41.28 & 4.41 & 0.06 & 3.55 & 0.30 & 3.70 & 0.15 & 1.2 & 17.65 & 16 & 9 & 10.22 & 160 & B2 \\
                        739 & 5:38:41.28 & -69:06:40.68 & 4.41 & 0.09 & 3.05 & 0.38 & 4.17 & 0.22 & 2.6 & 17.88 & 11 & 11 & 12.31 & 170 & B0 \\
                        2861 & 5:38:39.12 & -69:05:06.36 & 4.41 & 0.08 & 3.35 & 0.29 & 4.04 & 0.19 & 1.2 & 16.81 & 15 & 11 & 10.27 & 190 & B1 \\
                        1977 & 5:38:49.272 & -69:05:55.32 & 4.41 & 0.06 & 3.25 & 0.20 & 3.98 & 0.14 & 1.8 & 17.57 & 12 & 10 & 12.03 & 130 & B0 \\
                        762 & 5:38:49.728 & -69:06:43.2 & 4.40 & 0.01 & 4.18 & 0.08 & 5.66 & 0.04 & 1.6 & 13.10 & 50 & 37 & 3.37 & 150 & B0 \\
                        1279 & 5:38:45.72 & -69:06:22.68 & 4.40 & 0.01 & 4.38 & 0.05 & 5.99 & 0.05 & 1.8 & 12.48 & 81 & 57 & 2.52 & 140 & O9 \\
                        1951 & 5:38:50.28 & -69:06:04.32 & 4.40 & 0.07 & 4.38 & 0.05 & 4.85 & 0.16 & 1.4 & 14.92 & 81 & 17 & 2.52 & 330 & B0 \\
                        1689 & 5:38:36 & -69:06:09.36 & 4.40 & 0.08 & 4.38 & 0.05 & 5.81 & 0.19 & 1.8 & 12.92 & 81 & 46 & 2.52 & 260 & B0 \\
                        605 & 5:38:44.952 & -69:06:44.64 & 4.36 & 0.10 & 2.54 & 0.26 & 3.76 & 0.24 & 1.4 & 17.41 & 7 & 8 & 7.25 & 180 & B0 \\
                        2190 & 5:38:45.6 & -69:05:47.76 & 4.36 & 0.06 & 4.34 & 0.05 & 5.79 & 0.15 & 1.6 & 12.55 & 83 & 43 & 2.31 & 130 & B1 \\
                        2438 & 5:38:39.096 & -69:05:19.68 & 4.34 & 0.07 & 2.86 & 0.19 & 3.86 & 0.18 & 1.0 & 16.64 & 8 & 9 & 20.12 & 100 & B3 \\
                        2572 & 5:38:43.944 & -69:05:19.68 & 4.34 & 0.09 & 3.26 & 0.28 & 4.05 & 0.22 & 1.4 & 16.57 & 10 & 9 & 20.05 & 150 & O9 \\
                        1324 & 5:38:44.448 & -69:06:18.72 & 4.34 & 0.12 & 3.16 & 0.43 & 3.49 & 0.30 & 1.2 & 17.76 & 9 & 7 & 22.51 & 180 & B0 \\
                        1682 & 5:38:46.128 & -69:06:07.2 & 4.34 & 0.11 & 3.26 & 0.38 & 3.71 & 0.25 & 1.6 & 17.61 & 10 & 8 & 20.05 & 150 & B0 \\
                        1952 & 5:38:31.824 & -69:05:56.76 & 4.34 & 0.10 & 3.26 & 0.29 & 3.88 & 0.25 & 1.4 & 16.99 & 10 & 9 & 20.05 & 220 & B0 \\
                        2017 & 5:38:50.304 & -69:05:52.44 & 4.32 & 0.12 & 3.48 & 0.36 & 3.55 & 0.29 & 1.4 & 17.71 & 15 & 7 & 16.80 & 200 & B0 \\
                        1399 & 5:38:35.712 & -69:06:17.28 & 4.32 & 0.07 & 3.98 & 0.15 & 4.82 & 0.16 & 2.2 & 15.32 & 29 & 16 & 6.49 & 130 & B3 \\
                        322 & 5:38:43.08 & -69:06:53.64 & 4.32 & 0.08 & 2.98 & 0.33 & 3.58 & 0.20 & 1.6 & 17.82 & 9 & 7 & 24.45 & 220 & O9 \\
                \end{longtable}
        \end{landscape}



\end{document}